\documentclass[letter,12pt]{article}
\usepackage[utf8]{inputenc}
\usepackage{amssymb}
\usepackage{latexsym}
\usepackage{amsmath}
\usepackage{graphicx} 
\usepackage{multirow}
\usepackage{color}   
\usepackage{hyperref}
\usepackage{adjustbox}
\usepackage{ulem}
\usepackage{caption} 
\usepackage[left=2cm,right=2cm,top=2.5cm,bottom=2.5cm]{geometry}
\hypersetup{urlcolor=blue, colorlinks=true}
\usepackage[numbers,sort&compress]{natbib}

\title{Top squark signal significance enhancement by different Machine Learning Algorithms}

\author{
Jorge Fraga\footnote{Email: jf.fraga@uniandes.edu.co}, 
Ronald Rodriguez\footnote{Email: rrodriguez@uniandes.edu.co},
Jesus Solano \footnote{Email: ja.solano588@uniandes.edu.co}, \\
Juan Molano\footnote{Email: jp.molano56@uniandes.edu.co}, 
 and 
Carlos Ávila\footnote{Email: cavila@uniandes.edu.co} \vspace{3mm} \\
 \textit{Department of Physics, Universidad de los Andes, } \\
\textit{Bogotá - Colombia.}}

\date{\today \\}

\begin{document} 
\maketitle

\abstract{ 
A study of four different machine learning (ML) algorithms is performed to determine the most suitable  ML technique to disentangle a hypothetical supersymmetry signal from its corresponding Standard Model (SM) backgrounds and to establish their impact on signal significance.
The study focuses on the production of SUSY top squark pairs (stops), in the mass range of $500<m_{\tilde{t}_1}<800$ GeV, from proton-proton collisions with a center of mass energy of 13 TeV and an integrated luminosity of 140 fb$^{-1}$, emulating the data-taking conditions of the run II LHC accelerator. In particular, the semileptonic  channel is analyzed, corresponding to final states with a single isolated lepton (electron or muon), missing transverse energy, and four jets, with at least one tagged as $b$-jet. The challenging compressed spectra region is targeted, where the stop decays mainly into a $W$ boson, a $b$-jet, and a neutralino ($\tilde{t}_1\rightarrow W+b+\tilde{\chi}_1^0$), with a  mass gap between the stop and the neutralino of about 150 GeV.
The ML algorithms are chosen to cover different mathematical implementations and features in machine learning. We compare the performance of a logistic regression (LR), a Random Forest (RF), an XGBoost (XG), and a Neural Network (NN) algorithm. 
Our results indicate that XG and NN classifiers provide the highest improvements (over 17\%) in signal significance, 
when compared to a standard analysis method based on sequential requirements  of different kinematic variables. The improvement in signal significance provided by the NN increases up to 31\% for the highest stop mass considered in the present study (800 GeV). The RF algorithm  presents a smaller improvement that decreases with stop mass.  On the other hand, the LR algorithm shows the worst performance in signal significance which even does not compete with the results obtained by an optimized cut and count method.
}

\section{Introduction}

The popularity of Machine Learning (ML) algorithms have expanded to different areas of knowledge, leaning in the last decade to Deep Learning algorithms  \cite{learn,ML}. In the field of High Energy Physics (HEP), ML algorithms have been used in different experimental searches, for instance, in particle reconstruction and identification \cite{dnnLHC, dnnLHC2,anomaly,synchrotrons,trigger,calorimetry}, physics measurements \cite{dnnLHC3, dnnLHC4,mesons}, and fast data processing  \cite{griddnn1,fastdata}. 
The Large Hadron Collider (LHC) experiments have been applying different ML techniques  to analyze and extract relevant information from the large amount of data collected by each specific detector \cite{DLLHC}. 
Some of the most common approaches are neural networks, support vector machines, random forest, boosted decision trees (BDT), and others \cite{TMVA}. Many of these algorithms are mainly used for identifying and recording data related to the physical properties of particles, for example, in  particle path reconstruction based on the information collected by tracker detectors and jet identification in hadronic calorimeters \cite{MLjets}.
Additionally, ML algorithms are extensively involved in the optimization of data storage  and information flow in the computing GRID centers worldwide \cite{griddnn2,griddnn3}.
Moreover, the implementation of ML algorithms played an important role in the discovery of the Higgs Boson in 2012 \cite{higgs2012atlas,higgs2012cms}, where the analyses were performed using BDTs. 

In this study, four ML algorithms are implemented with the purpose of discriminating SM backgrounds from beyond Standard Model (BSM) hypothetical processes in order to maximize the significance. We compare this approach to traditional analyses based on event cut-and-count methodologies. 
To accomplish this goal, a case of study in the search for SUSY particles is proposed, focusing on the production of stop pairs in proton-proton collisions at 13 TeV in the LHC environment. 
Particularly, we look for stops with R-parity conservation with benchmark mass points along the so-called compressed scenario where the difference between the mass of the stop and the neutralino is smaller than the mass of the on-shell top quark \cite{comprimidos, compressed1,compressed2}. 

This article is structured as follows. Section \ref{sec:susy_search} contains a brief introduction to the current searches of SUSY stop particles in the LHC and introduces the ML algorithms used in the present investigation.
Section \ref{sec:samples} provides all the details related to the MC sample generation used for the current study and describes the two strategies whose performance is compared:  signal extraction using standard methodologies based on event cut-and-count and  signal extraction using ML algorithms. For the latest method, the kinematic features and the structure of the data sets used for training and evaluation are also specified.
Results of the performance of both strategies are presented in section \ref{sec:results}. Additionally, for the ML algorithms,  a comparative analysis is performed based on classification metrics and a significance optimization for each signal benchmark. 
A comparison of signal significance obtained by the four ML classification algorithms and the standard classical method is made in terms of the gain in significance obtained from each analysis. Finally, conclusions are drawn in section \ref{sec:conclusions}.

\section{Search for SUSY stop particles }
\label{sec:susy_search}

Despite its mathematical rigor and precise predictions, the SM of particle physics does not provide a complete description of nature. It still presents several shortcomings such as being unable to provide any explanation for neutrino masses \cite{neutrinos}, the nature of dark matter \cite{darkmatter} and dark energy \cite{darkenergy}, the strong CP violation \cite{cp}, among others. Additionally, The SM also runs into quadratic divergences from top quark loop corrections in the Higgs mass Feynman diagrams, known as the hierarchy problem \cite{hierarchy, hierarchy2,hierarchy3}.

Supersymmetry (SUSY) is one of the most popular hypothesis to search for physics BSM, since it works as an extension of the SM, requiring an additional symmetry between fermions and bosons \cite{susytheory,susytheory2,susytheory3}. Within SUSY, some variants assume different couplings and free parameters, which determine how this symmetry is broken and the mass spectra of the supersymmetric particles \cite{susystates,susybroken1,susybroken2,susybroken3,susybroken4,susybroken5,susybroken6, strategy1,strategy2}. A relevant particle when looking for signatures of SUSY is the top squark $\tilde{t}_1$ (stop), since its existence could imply a natural solution to the hierarchy problem \cite{stops}. As the stop is the super-partner of the top quark, its positive contribution to the Higgs mass corrections cancels out naturally the largest negative diverging contribution coming from the SM top quark. Additionally, in most SUSY models with R-parity conservation, the lightest supersymmetric particle is the neutralino $\tilde{\chi}_1^0$, which arises naturally as a strong candidate for dark matter, due to its large mass, weakly interactive nature, and extended lifetime \cite{lsp}.

\begin{figure}[h!]
\centering
\includegraphics[width=0.4\textwidth]{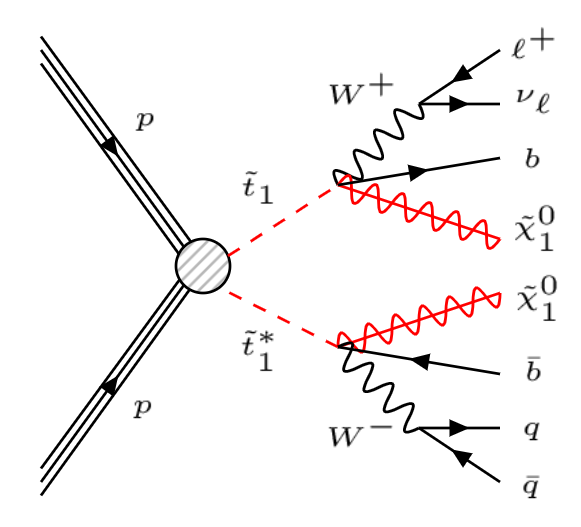}
\caption{A representative Feynman diagram for stop pair production with three body decay and semileptonic final states, with $\ell=e,\mu$.}
\label{fig:feynman}
\end{figure}

This study explores the production of pairs of SUSY stop particles from proton-proton collisions in a  compressed spectra scenario, where the mass difference between the top squark and the lightest neutralino is smaller than the top quark mass by about 20 GeV, causing the particles of the final state to have very low kinetic energy \cite{comprimidos,compressed1,compressed2}. 
In this case, each stop decays dominantly to an off-shell top quark and a neutralino, with the top quark subsequently decaying into an on-shell $W$ boson and a $b$-quark. The two neutralinos in the final state are usually produced in opposite directions, generating a very low missing transverse momentum in the event, increasing the experimental challenge to distinguish the SUSY signal from its SM backgrounds.
There are three final state topologies in the direct stop pair-production, depending on how each $W$ boson in the final state decays: the full hadronic, semileptonic, and dileptonic channels. This study focuses in the semileptonic channel, where one of the $W$ bosons decays into quarks and the second one into a charged lepton $\ell$ (electron or muon) and a neutrino ($ W \rightarrow \ell \nu $), corresponding to events with final states with a single isolated lepton,  missing transverse momentum, and four jets with at least one of them tagged as a $b$-jet. 
An important advantage of this specific topology is the reduction of QCD background due to the  requirement of  the isolated lepton \cite{cheng,macaluso}. A schematic Feynman diagram of this process is shown in Figure \ref{fig:feynman}. 

So far, searches by  CMS and ATLAS experiments for stops in the semileptonic channel have yielded null results and established exclusion limits for the stop mass up to about 1 TeV and for neutralino masses approximately below 600 GeV \cite{semileptonicatlas,semileptoniccms}. 

On the other hand, ML methods are by now a well-established framework used in most HEP experiments to improve the performance in the detection of particles, analyzing in principle all multiple correlations between their features 
\cite{dnnLHC, dnnLHC2, dnnLHC3, dnnLHC4}.
In the last decade, the improvement of ML algorithms implied, for instance, larger and more robust neural networks with many hidden layers and nodes.
These improvements have increased the capabilities to 
extract deeper and meaningful information from the combination of several features of the large amount of data stored by any specific experiment at the LHC \cite{dnn1,dnn2}.
The use of ML algorithms has been rapidly extended to the analysis of phenomenological and experimental studies, for example, in searches of new physics phenomena such as anomaly detection or the decay of hypothetical Supersymmetric particles \cite{dnn3,dnn4,dnn5,dnn6, dnn7, dnn8,dnn9}. ML algorithms have already been used in the search for stop production to discriminate final decay products from SM backgrounds \cite{dnn10}. In reference \cite{semileptonicatlas}, the ATLAS experiment reports an analysis for the production of pairs of stops in the semileptonic channel using a recurrent neural network to reduce jet multiplicity and a shallow neural network to perform the discrimination between SUSY signals and backgrounds. The present study expands and improves this type of analysis
with the implementation of neural networks with more hidden layers and neurons to improve the training, and by making a comparison with additional ML algorithms in order to determine the one with the highest performance in enhancing signal significance for stop masses up to 800 GeV via the three-body decay ($\tilde{t}\rightarrow b W \tilde{\chi}^{0}_{1}$). 

The effectiveness to discriminate SM backgrounds from stop decay processes is investigated for the following ML algorithms: a Logistic Regression \cite{logistic}, which basically estimates the conditional probability of a dichotomous variable by fitting the model parameter using maximum likelihood; a Random forest \cite{forest} classifier, which constructs many different decision trees that classify a sample by majority voting; a Gradient Boosting \cite{xgboost} (XGBoost) classifier, which is an ensemble method that sequentially improves a base decision tree by means of optimizing a loss function; and finally, a sequential neural network \cite{DNN}  classifier (NN), which includes multiple hidden layers and neurons to encode non-linear relationships between inputs and outputs. It is important to remark that the selection of these specific classifiers was aimed to have a wide range of models increasing in complexity. Note that the features behind the proposed classifiers are different which let us understand and choose the best approach to our task.  

\section{Event samples and search strategies }
\label{sec:samples}
\subsection{Event samples}
\label{subsec:event_samples}
MC samples of signal and background events are generated following similar criteria as previous experimental searches of production of stop pairs from proton collisions at the LHC. 
Samples are generated at leading order in the matrix elements for proton-proton collisions with a center of mass energy of 13 TeV.
For the hard collisions at parton level, we use Madgraph 5 \cite{madgraph,madgraph1}, and  Pythia 8 \cite{pythia8} for showering and hadronization simulation. Finally, the detector simulation has been implemented in Delphes \cite{delphes}, 
using the anti-$k_t$ clustering method and FastJet \cite{antikt, fastjet} for jet reconstruction, with a radius parameter of $R = 0.5$, where $R=\sqrt{\eta^2+\phi^2}$ with $\eta$ the pseudorapidity and $\phi$ the azimuthal angle. For b-tagging, $E_T^{miss}$, and lepton identification efficiency, we use the parameters established in Delphes to simulate the performance of the CMS detector \cite{cmsdetector}.

\subsubsection*{SUSY Signal}
The signal sample parameters are produced using the SoftSusy  package \cite{softsusy} based on the Minimal Supersymmetric Standard Model (MSSM). 
A large scan over the MSSM parameters is performed to look for signal signatures lying along the compressed mass spectrum. In our study we require that the difference in mass between the stop and the lightest neutralino to be around 150 GeV. We have selected four signal benchmark mass points taking into account the NNLO-NLL cross-sections \cite{crosssections}. Other parameters of the benchmarks selected are shown in Table \ref{tab:signalparameters}, where the notation \textit{s500} refers to the benchmark with stop mass of 500 GeV, and so on. This notation is used throughout this article to refer briefly to our signal benchmark points. The process depicted in Figure \ref{fig:feynman} is simulated with up to two additional jets at parton level in order to take into account effects from initial state radiation (ISR).

\begin{table}[h!]
    \centering
    \caption{MSSM signal parameters  for each benchmark in the present study.
    }
    \label{tab:signalparameters}
    \begin{tabular}{lcccc}
    \hline    \hline
     Parameters & \textit{s500} & \textit{s600} & \textit{s700} & \textit{s800}  \\
    \hline
$m(\tilde{t})$ [GeV] & 500 & 600 & 700 & 800 \\
$m(\tilde{\chi}^0_1)$ [GeV] & 350 & 450 & 550 & 650 \\
$\sigma$ (NNLO-NLL) [fb]  & 0.6090 & 0.2050 & 0.0783  & 0.0326 \\
$\Omega h^2$  & 2.94 & 5.34 & 5.97  &  8.74\\
$m_0$ [GeV] & 2450& 2600 & 2560 &2400 \\
$m_{1/2}$ [GeV] & 800& 950 & 1240 &1400 \\
$A_0$ [GeV] & -5950 & -6500 & -7200& -7300\\
$\tan \beta$  & 25 & 25 & 35 & 32\\
\hline     \hline
    \end{tabular}
\end{table}

\subsubsection*{Backgrounds}
For backgrounds, the processes shown in Table \ref{tab:backgrounds} are generated, which are the most representative for the channel under study, and consistent with other SUSY searches with a similar final state topology \cite{semileptonicatlas,semileptoniccms}. Since the $t\bar{t}$ process has a significant contribution, it has been divided into two channels, the semileptonic ($t\bar{t}$ 1L) and the dileptonic ($t\bar{t}$ 2L). 
The contribution from the hadronic channel  becomes negligible after demanding one lepton (electron or muon) in the final state, which is one of the pre-selection requirements discussed in the following section. 
Another significant background is $W+$jets, for which MC samples are produced after splitting it into three $H_T$ binned channels: 100-200, 200-400 and $H_T> 400$ GeV, where $H_T$ is the scalar sum of the transverse momentum of all hadrons in an event. 
The lower limit for $H_T$ of 100 GeV has been selected taking into account that there are at least four jets in the final state, whose scalar addition of transverse momentum is usually higher than this lower threshold.
Another important background is the single top production, for which we take into account the three main decay channels: $Wt$, $s$-channel, and $t$-channel. In the case of the latest channel, we have separated the $tbj$ and $tj$ final states production for convenience. We also consider  the $t\bar{t}+V$ production, where $V=W,Z$, and the diboson production channels: $WW$, $WZ$ and $ZZ$.  Up to two additional jets are allowed at generation level in  all background channels to allow for ISR jets.

In our study, we have required that all background processes which contain vector bosons $(W, Z)$ in their final states to decay to leptons ($\ell=e,\mu$) at parton level.
Therefore, a reduction in MC generated events is obtained by the smaller branching ratio of the decay to leptons. Scale factors have been applied to the background distributions to compensate the reduction of events with jets with respect to the inclusive processes.
Additionally, NLO k-factors \cite{factors} have been applied to normalize all background distributions.

\begin{table}[h!]
\centering
\caption{LO cross-sections for SM backgrounds at $\sqrt{s}=13$ TeV.}
\label{tab:backgrounds}
\begin{tabular}{llc}
\hline \hline
Background  & Sub-processes   &    $\sigma_{LO}\cdot BR [pb]$      \\ 
\hline
$t\bar{t}$ 1L  &  & 220.7    \\ 
$t\bar{t}$ 2L  &  & 40.17    \\ 
\hline
\multirow{3}{*}{$W+$jets}   & $H_T\in[100,200]$ GeV & 51.89   \\
                            & $H_T\in[200,400]$ GeV   & 17.68  \\
                            & $H_T\in[400,\infty]$ GeV   & 2.703  \\
\hline
\multirow{4}{*}{Single Top} & $Wt$  & 13.16    \\
                            & t-channel $(tbj)$ & 15.79     \\
                            & t-channel $(tj)$ & 5.166     \\
                            & s-channel $(tb)$  & 0.8814   \\ 
\hline
\multirow{2}{*}{$t\bar{t}+V$}   & $t\bar{t}+W$  & 0.3455   \\
                                & $t\bar{t}+Z$  & 0.5856  \\
\hline
\multirow{3}{*}{Diboson}    & $WW$  & 9.849 \\
                            & $WZ$  & 4.253 \\
                            & $ZZ$  & 3.752 \\
\hline \hline
\end{tabular}
\end{table}

\subsection{Event pre-selection for ML training}
Following the expected final states from the representative Feynman diagram of Figure \ref{fig:feynman}, we select events with only one lepton ($\ell=e,\mu$) with $p_{T}>$ 25 GeV and $|\eta|<2.5$  to remain within the detector's acceptance and to take into account the plateau trigger efficiency of CMS and ATLAS experiments for very well reconstructed leptons. Additionally, these selections reduce the hadronic $t\bar{t}$ and QCD multijet backgrounds. We require at least four reconstructed jets with $p_{T}>$ 25 GeV and $|\eta|<2.5$, and at least one of these jets to be $b$-tagged. Following the analysis strategies reported in
\cite{semileptonicatlas} and \cite{semileptoniccms}, to reduce QCD multijet background events and to ensure an isolated missing transverse momentum vector ($\vec{p}_T^{\; miss}$), we require that the azimuthal angular difference between the two leading jets and  $\vec{p}_T^{\; miss}$ to be greater than 0.4.

After making these selections, the efficiency to accept $W$+jets background events is very low, as can be observed in the event flow supplied in Table \ref{tab:eventflowb}. Note that the initial number of events of $W$+jets is more than four times the main background $t\bar{t}+1$L. Additionally, the $W$+jets background has a large cross-section (see Table \ref{tab:backgrounds}), implying a large number of events to be produced in order to satisfy the requirement of 140 fb$^{-1}$, to closely simulate the statistics collected by LHC experiments in Run II. 

On the other hand, we look for selections that enhance the discriminating capacity of the training features and, at the same time, keep the number of samples feasible for training. 
For this purpose, we optimize a selection in the transverse mass of the lepton and the missing transverse momentum,
\begin{equation}
\label{eq:mtl}
    m_T^{lep}=\sqrt{2\, p_T^{\ell} E_T^{miss}(1-\cos{\Delta \phi(p_T^{miss},p_T^\ell))}}.
\end{equation}
where $p_T^{\ell}$ is the transverse momentum of the lepton, $E_T^{miss}=\lVert\vec{p}_T^{\; miss} \rVert$ is the magnitude of the missing transverse momentum and $\Delta \phi(p_T^{miss},p_T^\ell)$ is the difference in azimuthal angles of the the corresponding transverse momentum vectors.
This selection reduces backgrounds such as $W$+jets and $t\bar{t}$ semileptonic since it reconstructs indirectly the $W$ boson from the final states. Likewise, we optimize a selection on $E_T^{miss}$ that comes mainly from the two energetic neutralinos in the final states of the signal events.

\begin{figure}[h!]
    \centering
    \includegraphics[scale=0.4]{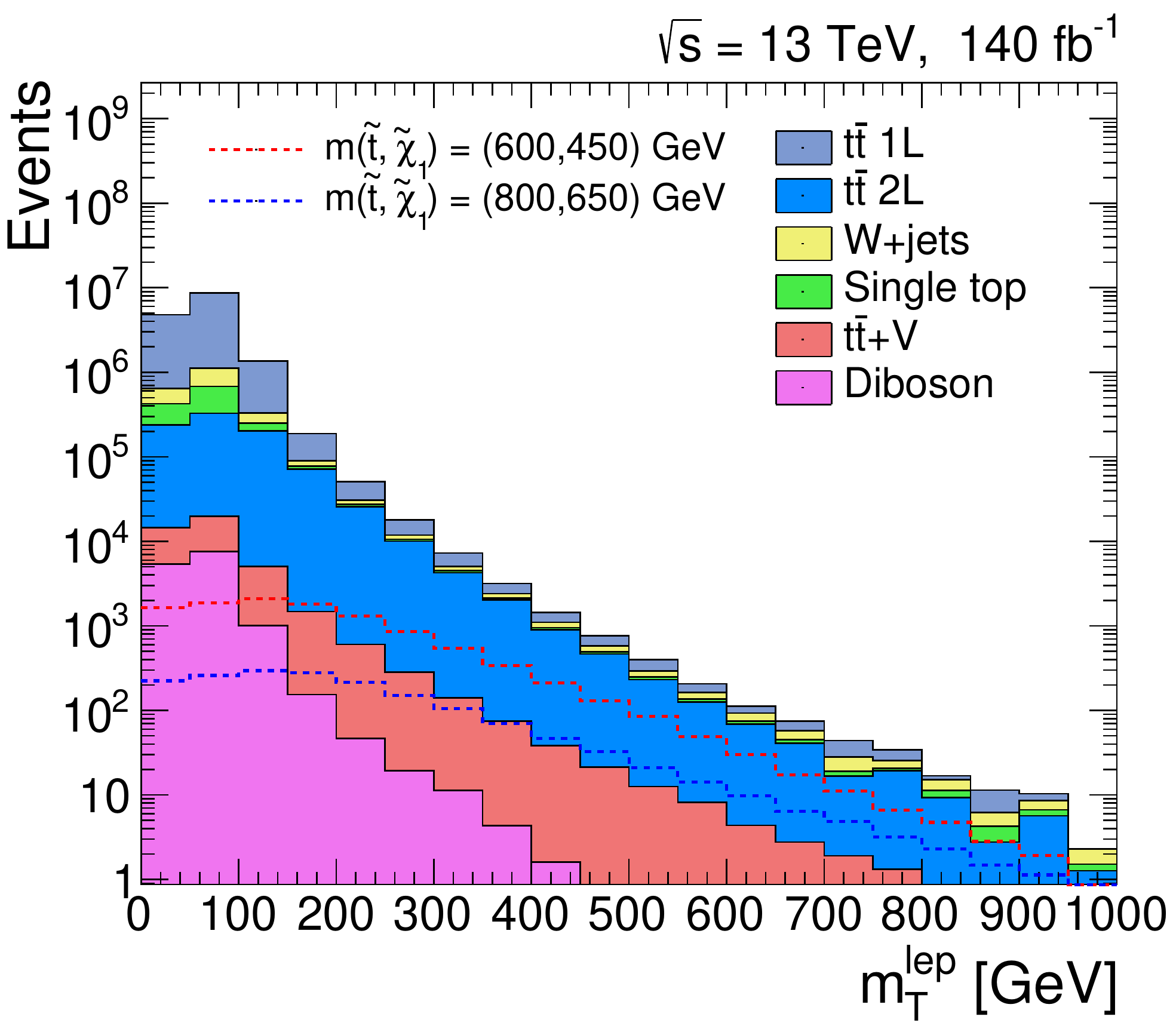}
    \includegraphics[scale=0.4]{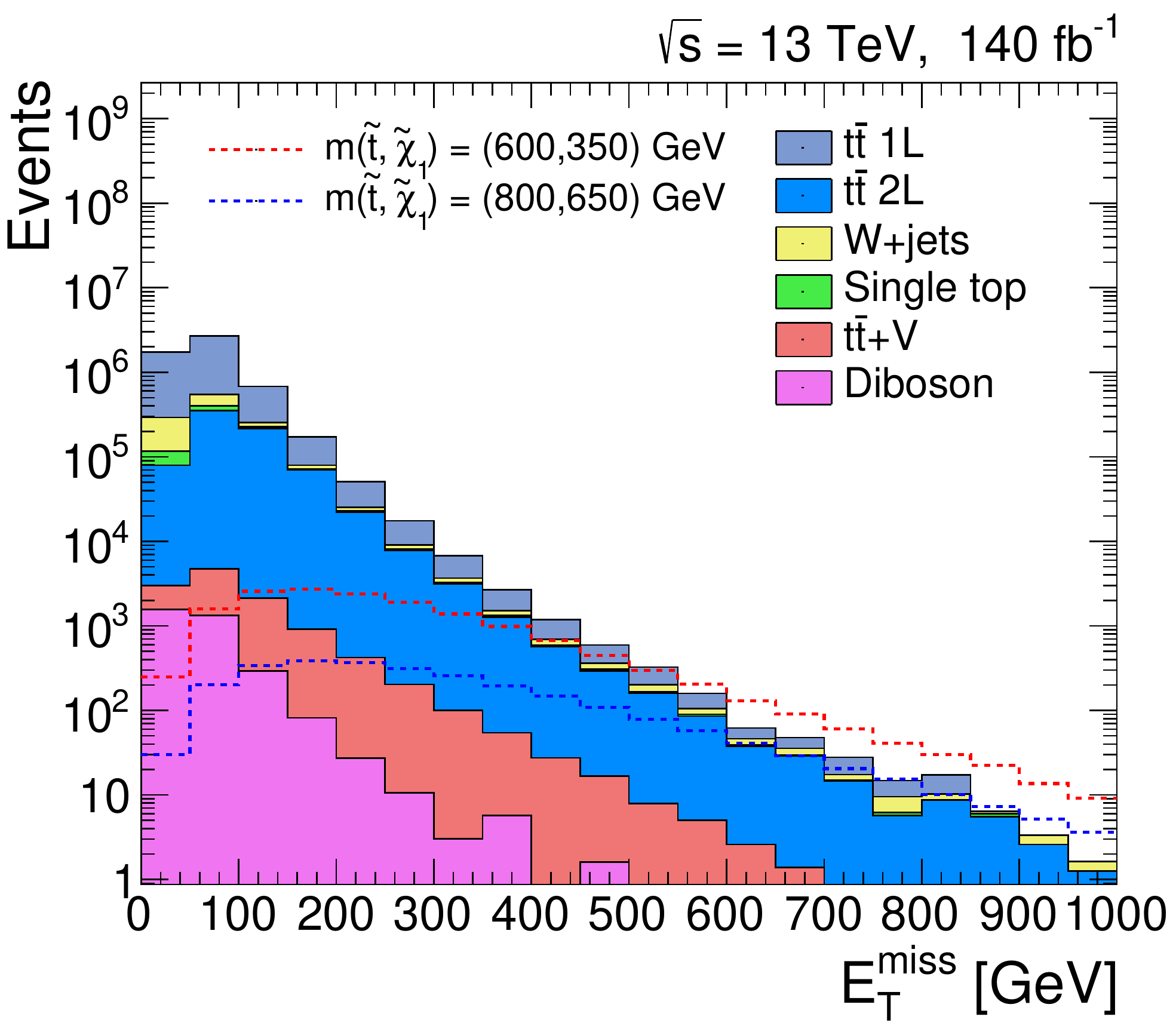}
    \caption{Distributions for transverse mass $m_T^{lep}$ (left panel) before the 90 GeV selection and $E_T^{miss}$ (right panel) after the $m_T^{lep}>90$ GeV selection. Two benchmark signals are displayed for comparison.}
    \label{fig:mtlmet}
\end{figure}

Considering the limitation in our computational capacity, we have found that selections of $m_T^{lep}>90$ GeV and $E_T^{miss}>90$ GeV allow us to generate a maximum of 400000 events for the ML training, divided equally into 50\% events for signal and 50\% for backgrounds.
The left panel of Figure \ref{fig:mtlmet} displays the  $m_T^{{lep}}$ distribution after the basic topological selections mentioned above. The right panel of Figure \ref{fig:mtlmet} indicates the $E_T^{miss}$ event distribution after the 90 GeV selection on $m_T^{{lep}}$. A summary of the event pre-selections is presented in Table \ref{tab:preselections}. Event flows for backgrounds and the signal benchmarks, after each pre-selection requirement,   are provided in Tables \ref{tab:eventflowb} and \ref{tab:eventflows}, respectively.

Finally, for evaluation, we have generated 140 fb$^{-1}$ for all  backgrounds except for $W$+jets background, where we have generated events only for an equivalent integrated luminosity of 86.9 fb$^{-1}$ (given our computational limitations) and then scale it to 140 fb$^{-1}$.  This scaling affects the uncertainty of the final background estimation. For the signal benchmarks, in order to have smooth distributions, we have generated higher statistics and then scale them to the same luminosity of background events.

\begin{table}[h!]
\centering
\caption{\label{tab:preselections} General pre-selection requirements. }
\begin{tabular}{ll}
\hline\hline
$N(\ell)$ ($\ell=e,\mu$)  		& =1 ($p_T>25$ GeV, $|\eta|<2.5$) \\
$N(j)$				    & $\geq 4$ ($p_T>25$ GeV, $|\eta|<2.5$)		\\
$N(\text{b-tag})$	    & $\geq 1$		\\
$\min \Delta\phi(j_{1,2},\vec{p}_T^{\; miss})$ 	& $>0.4$  	\\
\hline
$m_T^{{lep}}$			& $>90$ GeV \\
$E_T^{miss}$ 			& $>90$ GeV \\
\hline\hline
\end{tabular}
\end{table}

\begin{table}[ht!]
\centering
\caption{\label{tab:eventflowb}Event flow for backgrounds at a luminosity of 140 fb$^{-1}$.}
\resizebox{0.99\textwidth}{!}{
\begin{tabular}{lrrrrrrr}
\hline \hline 
       & \multicolumn{6}{c}{Backgrounds} \\ \cline{2-8} 
   & \multicolumn{1}{c}{$t\bar{t}$ 1L} & \multicolumn{1}{c}{$t\bar{t}$ 2L}   & \multicolumn{1}{c}{$W+$ jets} & \multicolumn{1}{c}{Single Top} & \multicolumn{1}{c}{$t\bar{t}+V$} & \multicolumn{1}{c}{Diboson} & \multicolumn{1}{c}{Total} \\ \hline
Initial Events                            & 45437200 & 8269769 & 36395000 & 5649060 & 190500 & 3566700 & 99508229 \\
$N(\ell=e,\mu)=1$,                        & 29025400 & 3737550 & 23455900  & 3924690 & 54862  & 1609600 & 61808002 \\
$N(j)\geq4$                               & 18117600 & 1270150  & 2765200  & 911986  & 44499  & 77820  & 23187255  \\
$N(\text{b-tag})\geq1$                    & 15729000  & 1112230  & 942017   & 743815   & 38818  & 18444   & 18584324  \\
$\min \Delta\phi(j_{1,2},\vec{p}_T^{\; miss})>0.4$ & 14000300  & 989968  & 808791   & 657935  & 34318  & 16262   & 16507574   \\
$m_T^{{lep}}>90$ GeV                      & 2064180   & 370894   & 153778    & 98123   & 8281   & 2072    & 2697328   \\
$E_T^{miss}>90$ GeV                       & 389030   & 193888   & 20444     & 18842    & 4023   & 349     & 626576   
\\ \hline \hline 
\end{tabular} 
}
\end{table}

\begin{table}[ht!]
\centering
\caption{\label{tab:eventflows}Event flow for signal benchmarks with 50000 simulated events and scaled to a luminosity of 140 fb$^{-1}$.}
\begin{tabular}{lrrrr}
\hline \hline 
 & \multicolumn{4}{c}{Signal}                                                  \\ \cline{2-5} 
    & \textit{s500}        & \textit{s600}           & \textit{s700} & \textit{s800}        \\ 
\hline
Initial Events                            & 85260 & 28700 & 10962 & 4564 \\
$N(\ell=e,\mu)=1$,                        & 53691 & 17961 & 6780  & 2790 \\
$N(j)\geq4$                               & 43076 & 14548 & 5534  & 2281 \\
$N(\text{b-tag})\geq1$                    & 37431 & 12641 & 4812  & 1982 \\
$\min \Delta\phi(j_{1,2},\vec{p}_T^{\; miss})>0.4$ & 34319 & 11637 & 4441  & 1835 \\
$m_T^{{lep}}>90$ GeV                      & 22845 & 7904  & 3104  & 1315  \\
$E_T^{miss}>90$ GeV                       & 20483 & 7203  & 2867  & 1225  \\ \hline\hline 
\end{tabular}
\end{table}

\newpage
\subsection{Signal extraction with a standard cut-and-count method}
\label{subsec:cut_count}

We use $E_T^{miss}$ as a critical variable to look for any signature of stop production, since large $E_T^{miss}$ is expected from neutralinos that cannot be detected in the final states (see Figure \ref{fig:feynman}). As a consequence, we expect signal events with large values of $E_T^{miss}$ compared to  background processes. This can be verified in the left panel of Figure \ref{fig:mtlmet} for the \textit{s600} and \textit{s800} benchmark distributions, where the last bin that contains the overflow tail events is enhanced.
The strategy we are following is to make convenient selections on other kinematic variables to boost the significance in the final distribution of $E_T^{miss}$, and afterwards make particular selections in $E_T^{miss}$ to maximize each benchmark significance.

An accurate way to determine the statistical significance of a signal $S$ standing on top of a background  $B$, which has an uncertainty $\sigma_B$, is by taking a profile likelihood ratio of a background only hypothesis which, for the case of $S$ and $B$ being gaussianly distributed, can be approximated as \cite{semileptonicatlas, significance1, significance2,significance3}:
\begin{equation}
\label{eq:sig}
    \text{Sig}=\sqrt{2}\Bigg\{(S+B)\ln\left[\frac{(S+B)(B+\sigma_B^2)}{B^2+(S+B)\sigma_B^2}\right]- \frac{B^2}{\sigma_B^2}\ln\left[1+\frac{\sigma_B^2 S}{B(B+\sigma_B^2)}\right]\Bigg\}^{1/2},
\end{equation}

where $S$  and $B$ are, respectively, the signal and background events remaining after applying the event selection criteria. The  background fluctuations could come from statistical and systematic uncertainties. Sources of the later are related to the total contributions of theoretical and experimental uncertainties from the MC simulators to get the cross-sections and the particle reconstruction efficiencies, respectively \cite{uncertainties}. Other sources  are related with the jet energy scale (JES), jet energy resolution (JER), $E_T^{miss}$, lepton and b-tagging identification \cite{uncertainties2,uncertainties3,uncertainties4}. We  take a conservative approach and consider that the contribution of systematic uncertainties are about 50$\%$ of their statistical counterpart, which are over-estimated as compared to the values reported by the CMS and ATLAS experiments \cite{semileptonicatlas,semileptoniccms}. With this assumption, $\sigma_B^2=1.25B$. 

Once all pre-selections are implemented, we proceed to optimize the significance of signal and background events in several distributions of kinematic variables. 
The most representative of these distributions  (normalized to unity) are shown in Figure \ref{fig:features}, which correspond to the features chosen for ML training, since they exhibit some distinctive traits in signal as compared to overall background. To generate these plots, we have stacked every background in one single histogram, and selected the \textit{s800}  signal for comparison.

Optimization has been performed in three most significant variables: 1)  The transverse mass produced by $\vec{p}_T^{\; miss}$ and the lepton ($m_T^{lep}$); 2) The azimuthal angular difference between $\vec{p}_T^{\; miss}$ and the $b$-jet transverse momentum ($\Delta \phi(\vec{p}_T^{\; miss},p_T^{b\text{-jet}})$); and 3) The variable $R_M$ that is defined as the ratio between $E_T^{miss}$ and the transverse momentum of the leading jet $p_T(j_0)$ \cite{liantao}, 
\begin{equation}
\label{eq:rm}
    R_M=\frac{E_T^{miss}}{p_T(j_0)}.
\end{equation}

\subsection{Signal extraction with ML}
\label{subsec:ML}

In order to discriminate signal from SM backgrounds in \textit{stop} decay processes, a more complex approach based on ML techniques is performed. The objective is to design a statistical method capable of learning from several traits of backgrounds and signal samples in the multi-dimensional space defined by  the features considered for this analysis. We  address this problem by using a binary classifier that categorizes between signal and background classes. A second possible approach is to use a multi-class algorithm to also identify the different background sources, however, for the present analysis we combine all backgrounds in one single class (according to their cross section contributions), since our main goal is to differentiate a specific SUSY benchmark signal from the overall background.

For our classification task, we train and evaluate a set of four ML algorithms that have different mathematical foundations. 
First, a logistic regression (LR) classifier \cite{logistic} that is basically a statistical model based on regression analysis in which a dependent binary variable is fitted using a logistic model \cite{logistic2}. Second, we have chosen the Random Forest (RF) \cite{forest} and the Gradient Boosting Classifiers \cite{xgboost}, which are sets of ensemble methods used to improve the performance of single classifiers \cite{esemble}. On this regard, the Random Forest is fundamentally a bagging ensemble of decision trees \cite{forest2}, while the Optimized Gradient Boosting Classifier (XG) is a meta-model that iterates over a set of weak classifiers \cite{xgboost2}. Finally, we propose a feed-forward Neural Network \cite{DNN} intended to build a decision boundary after stacking several hidden layers of non-linear relationships. 
It is important to remark that the difference in mathematical nature of the proposed classifiers let us grasp the traits of the latent space inherent to signal and background distributions. 
Additionally, with the aim of covering a wider range of ML algorithms, we also attempted to perform the signal extraction with a Support Vector Machine (SVM) classifier. However, as the complexity of SVM scales at least quadratically with the number of training events~\cite{scikit,svm}, we studied this classifier with a low number of training events, verifying that the computational time also scales quadratically with the training events. 
Additionally, significance values obtained with the SVM showed to be around 5\% better than the LR technique but lower than the RF and XG algorithms, around 2\% and 5\%,  respectively.
We have not included this approach in our study due to its practical limitations and its performance as compared to other ML algorithms in the preliminary tests performed. 

Regarding the details for each proposed classifier, for the Logistic Regression we apply a logit regression model with an optimized solver that uses a coordinate descent algorithm with regularization, which adds a squared penalization to model coefficients \cite{regular}. We use regularization in order to reduce the over-fitting and improve numerical stability during the training phase. Other parameters like the tolerance and the inverse of regularization strength were optimized with no improvement in significance. 

For the Random Forest classifier \cite{forest}, we propose a forest with 50 trees, and a Gini Impurity metric for splits and bootstrap \cite{gini1}. The number of trees  was optimized using a grid-search over the number of estimators in the model in the range of [50,100] trees. Furthermore, we use the Gini impurity metric because it is less intensive than other split criteria when splitting nodes in decision trees. 

For the Optimized Gradient Boosting, we use an ensemble of 100 trees boosted using the tree booster (\textit{gbtree}). We chose a tree booster because we are iterating over weak decision trees \cite{xgboost}. The number of the trees in this model was selected after a grid-search in the range of [50,500] trees, where the significance was optimized. In that sense, each decision tree has 100 estimators. Additionally, the $\eta$ parameter has been optimized in the range [0.01,0.30], getting a value of 0.30. Note that a XGBoost is also an ensemble classifier like the Random Forest, however, the XGBoost is based on boosting weak learners (shallow decision trees) that use gradient descent algorithms over the error metric. On the other hand, the random forest is a bagging of fully grown decision trees which are ensembled \cite{esemble}.

Finally, for the NN we propose a fully-connected neural network with \textit{sigmoid} activation in the last layer and rectified linear units (\textit{ReLu}) activation in previous layers \cite{rectified}. We optimized the activation for the intermediate layers between \textit{tanh} and \textit{sigmoid}. The \textit{sigmoid} is commonly used for solving binary classification problems. 
Note that a \textit{sigmoidal} behavior is needed as we aim to warranty the final layer to be insensitive to small changes in the saturation states. We performed a hyper-parameter optimization on the learning rate, number of layers and nodes per layer. 
The learning rate was optimized in the range [$10^{-7}$, $10^{-2}$], obtaining an optimal value of $10^{-4}$.  Layers were optimized in the range [3,10] obtaining an optimal value of 4 layers. Nodes for the three intermediate \textit{ReLu} layers were optimized in the range [4,128] obtaining optimal values of 32, 16 and 8 nodes for the first, second and third layers, respectively. This optimization was carried out by comparing the validation accuracy for different configurations. 
In the case of the NN, we normalized the feature distributions fed to the neural network in order to remove biases generated by any difference in the features dimensionality. Moreover, the feature normalization is needed to equally distribute the importance of all variables in the training phase. We tested different normalization techniques: Standardizing, Power Transform normalization and Quantile normalization \cite{scikit}. The best improvement in significance was obtained when normalizing the features through a Quantile transformer algorithm  with gaussian output. 

\subsubsection*{Features for ML  Training}

\begin{figure}[ht!]
    \centering
\includegraphics[width=0.32\textwidth]{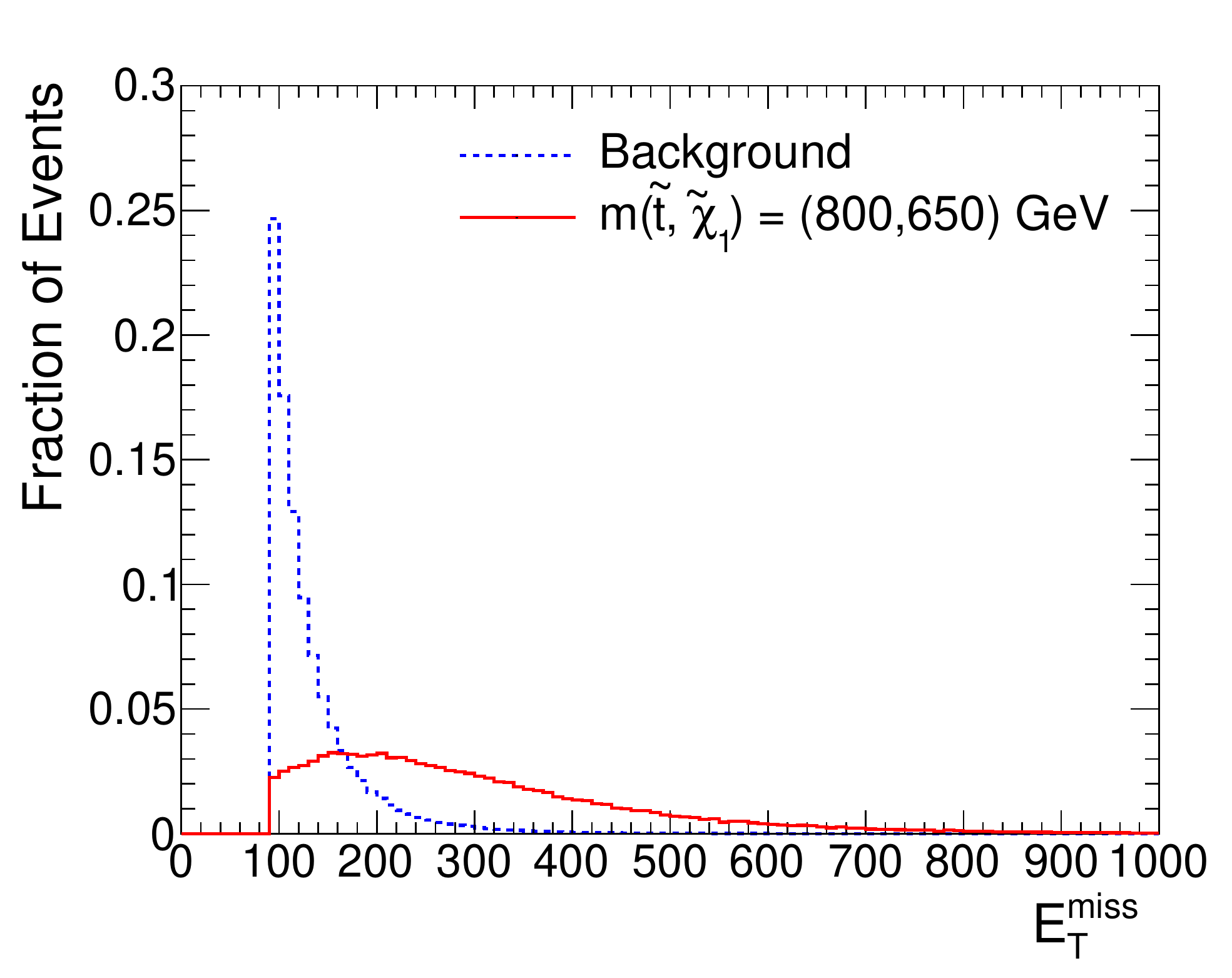}
\includegraphics[width=0.32\textwidth]{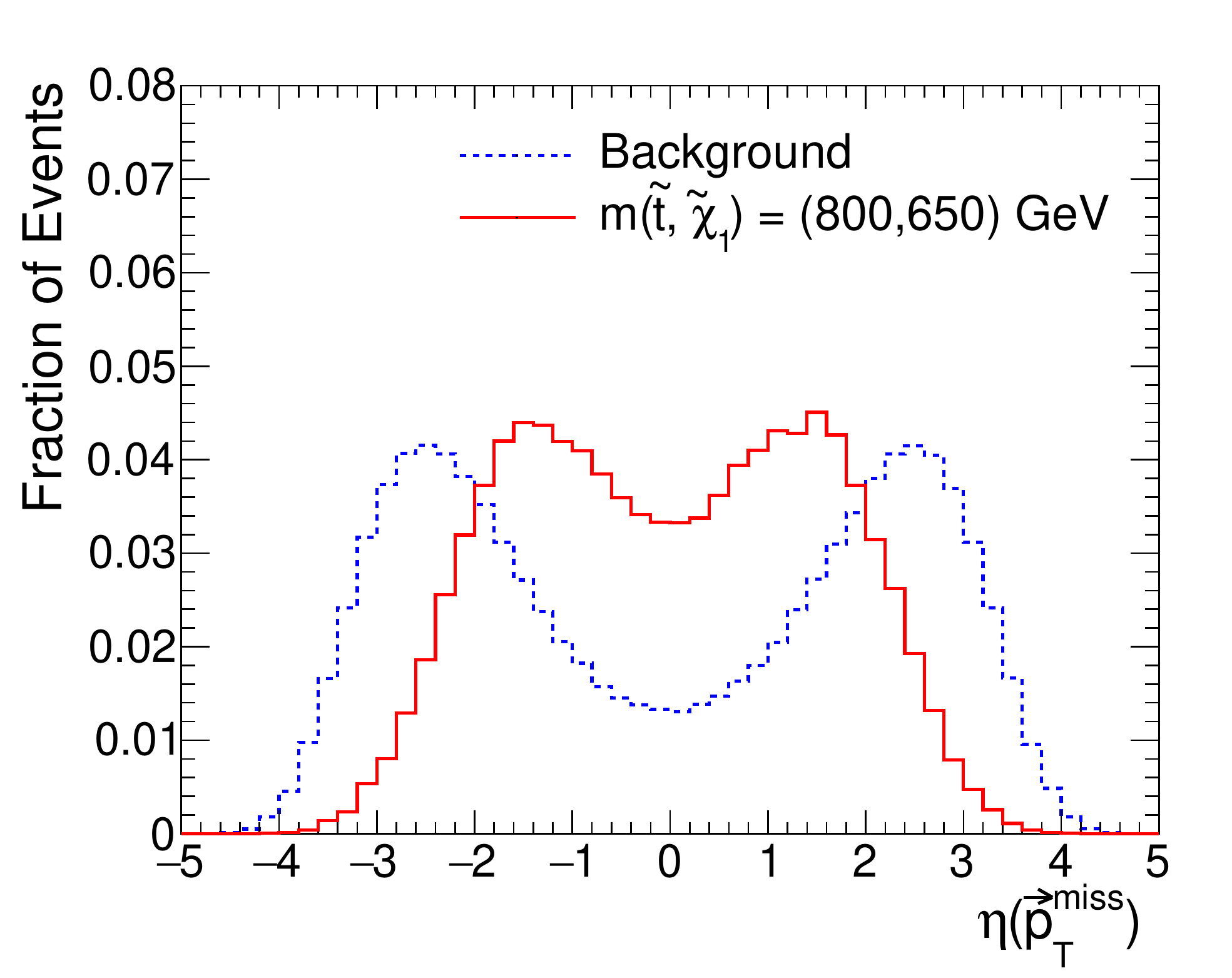}
\includegraphics[width=0.32\textwidth]{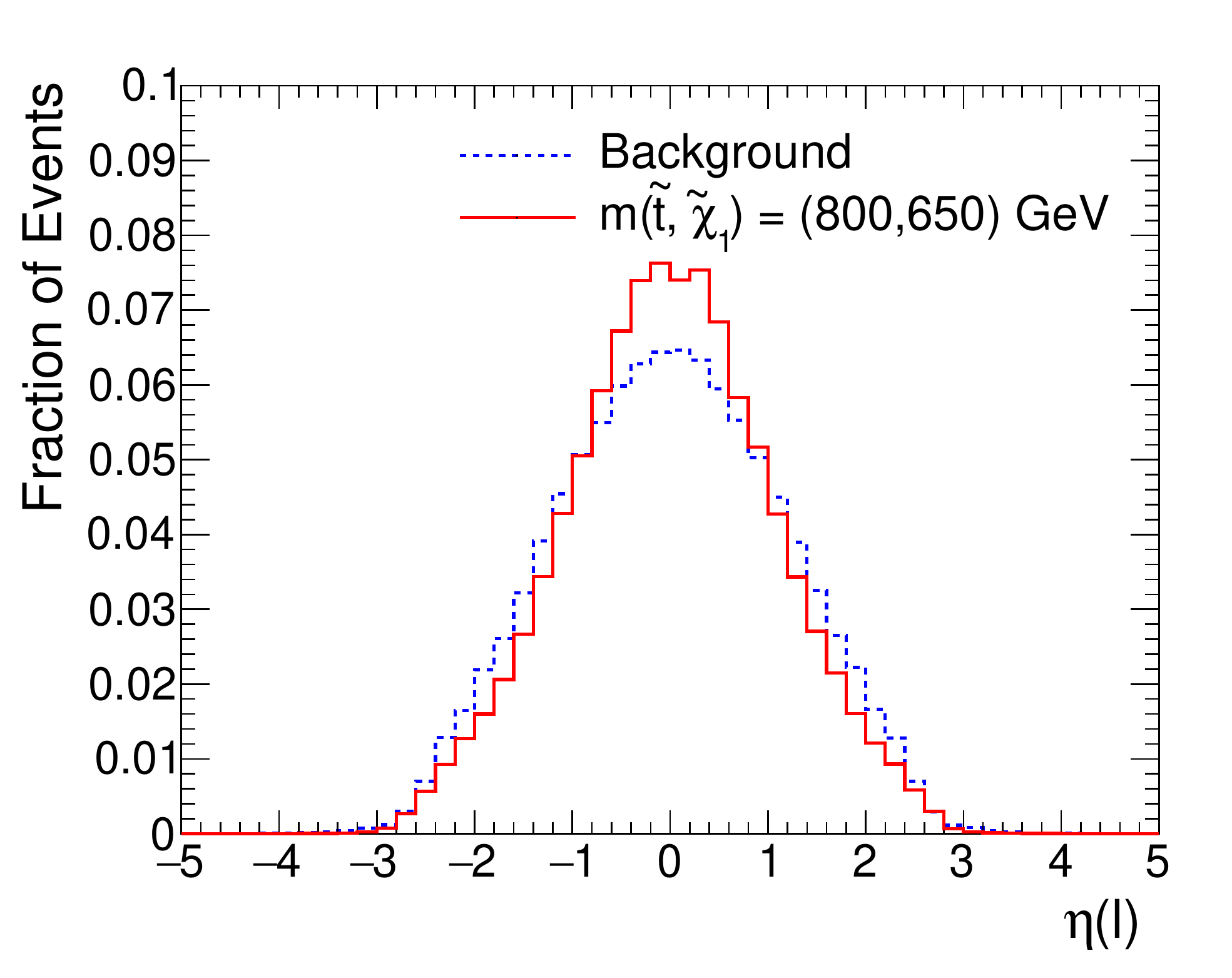}
\includegraphics[width=0.32\textwidth]{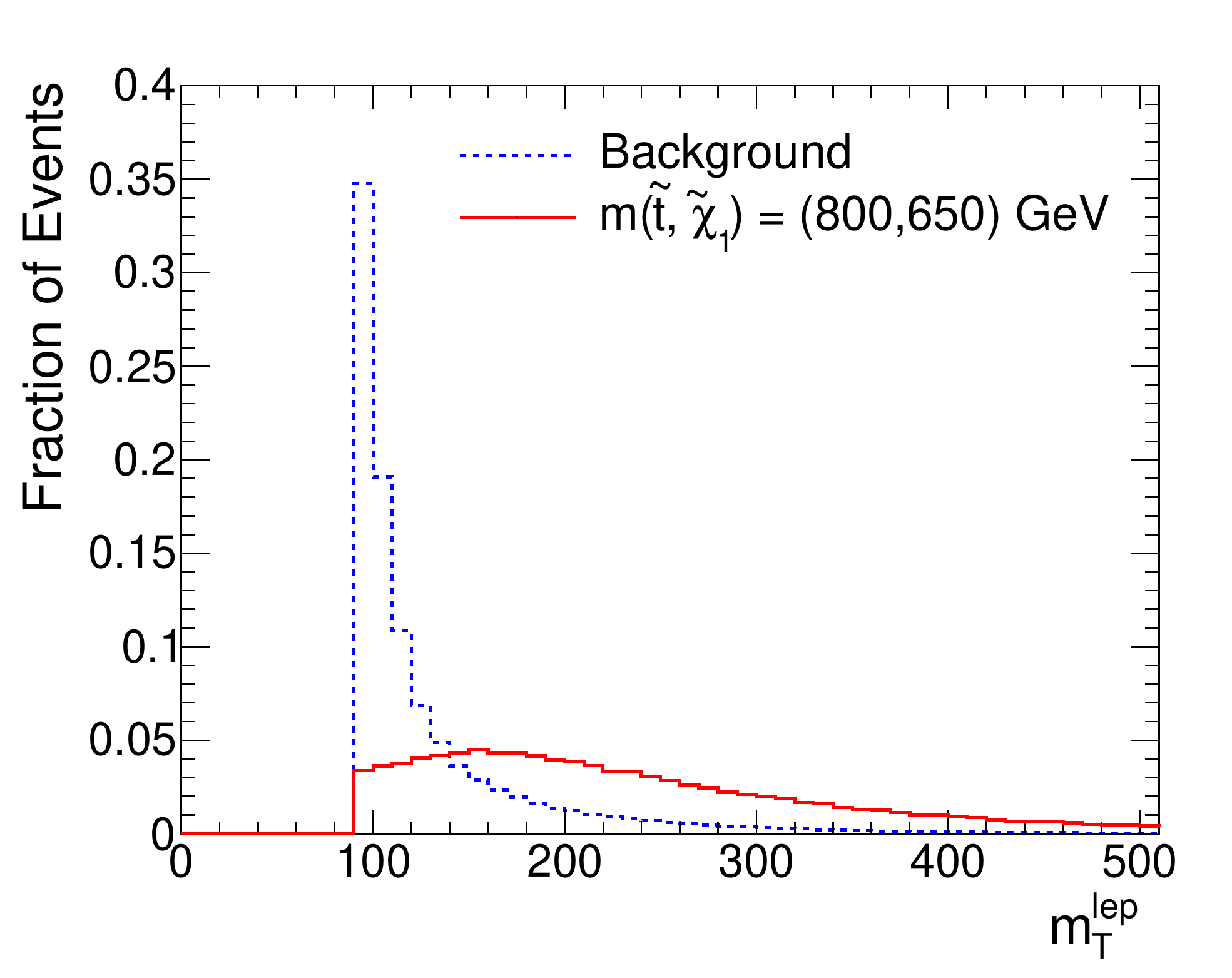}
\includegraphics[width=0.32\textwidth]{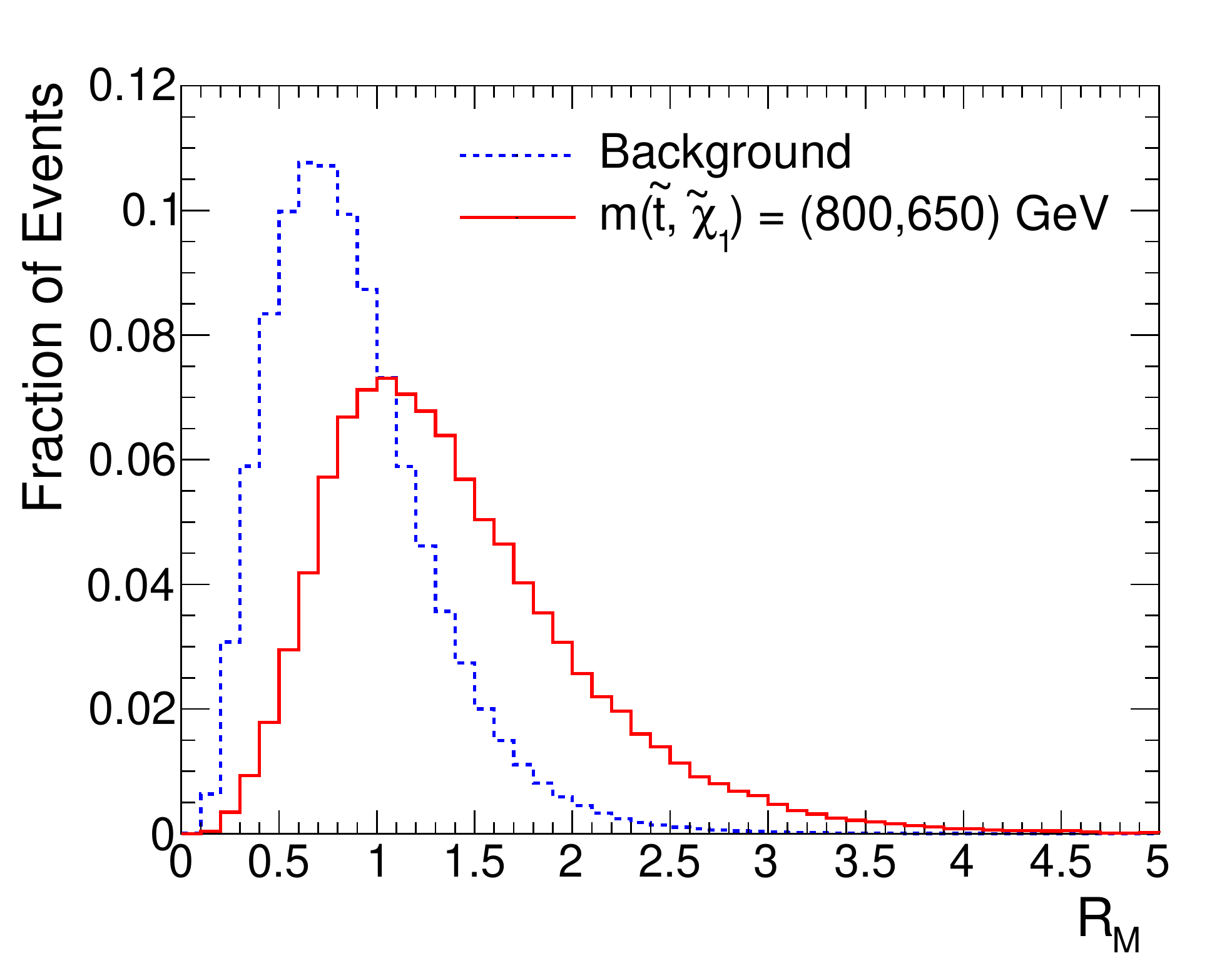}
\includegraphics[width=0.32\textwidth]{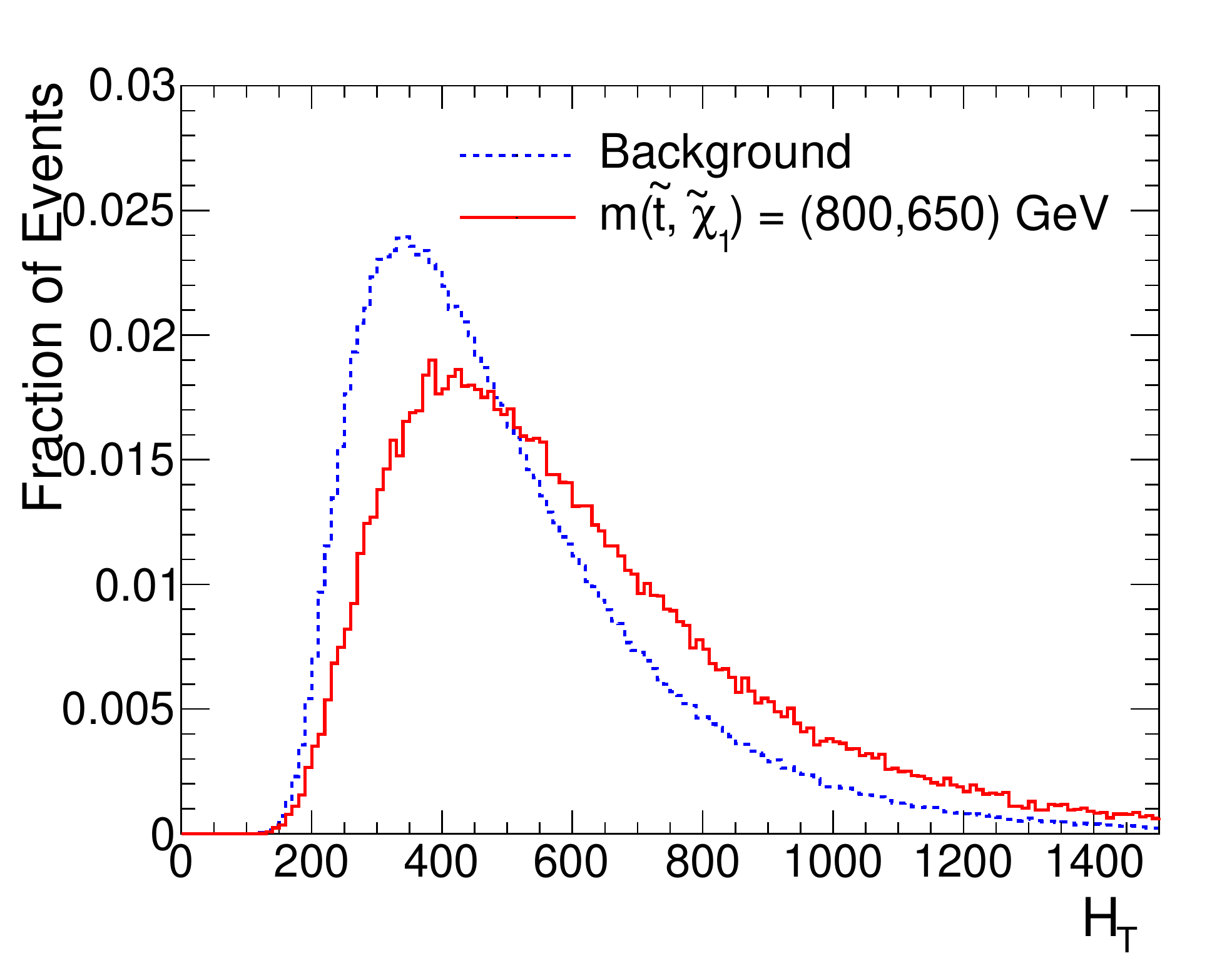}
\includegraphics[width=0.32\textwidth]{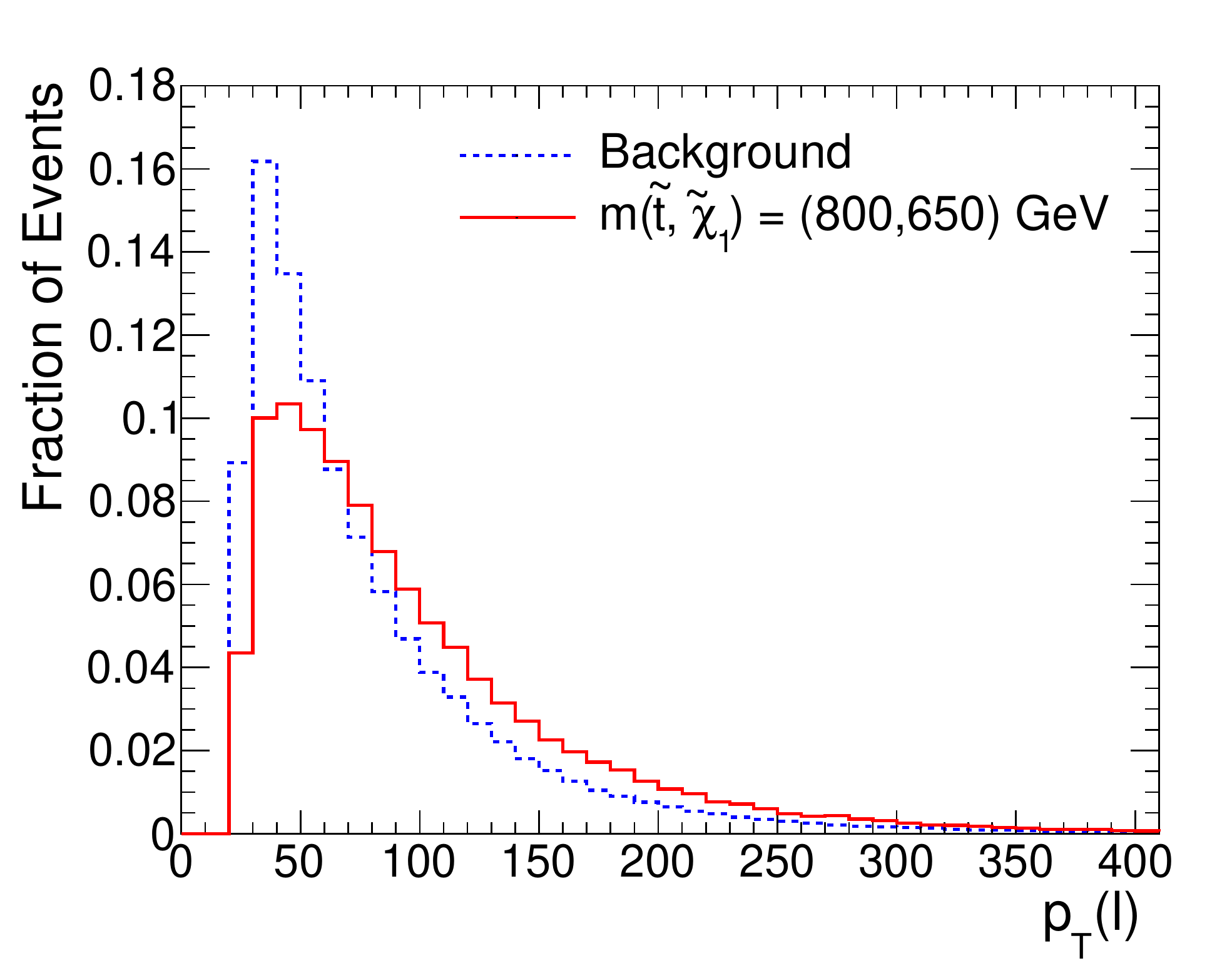}
\includegraphics[width=0.32\textwidth]{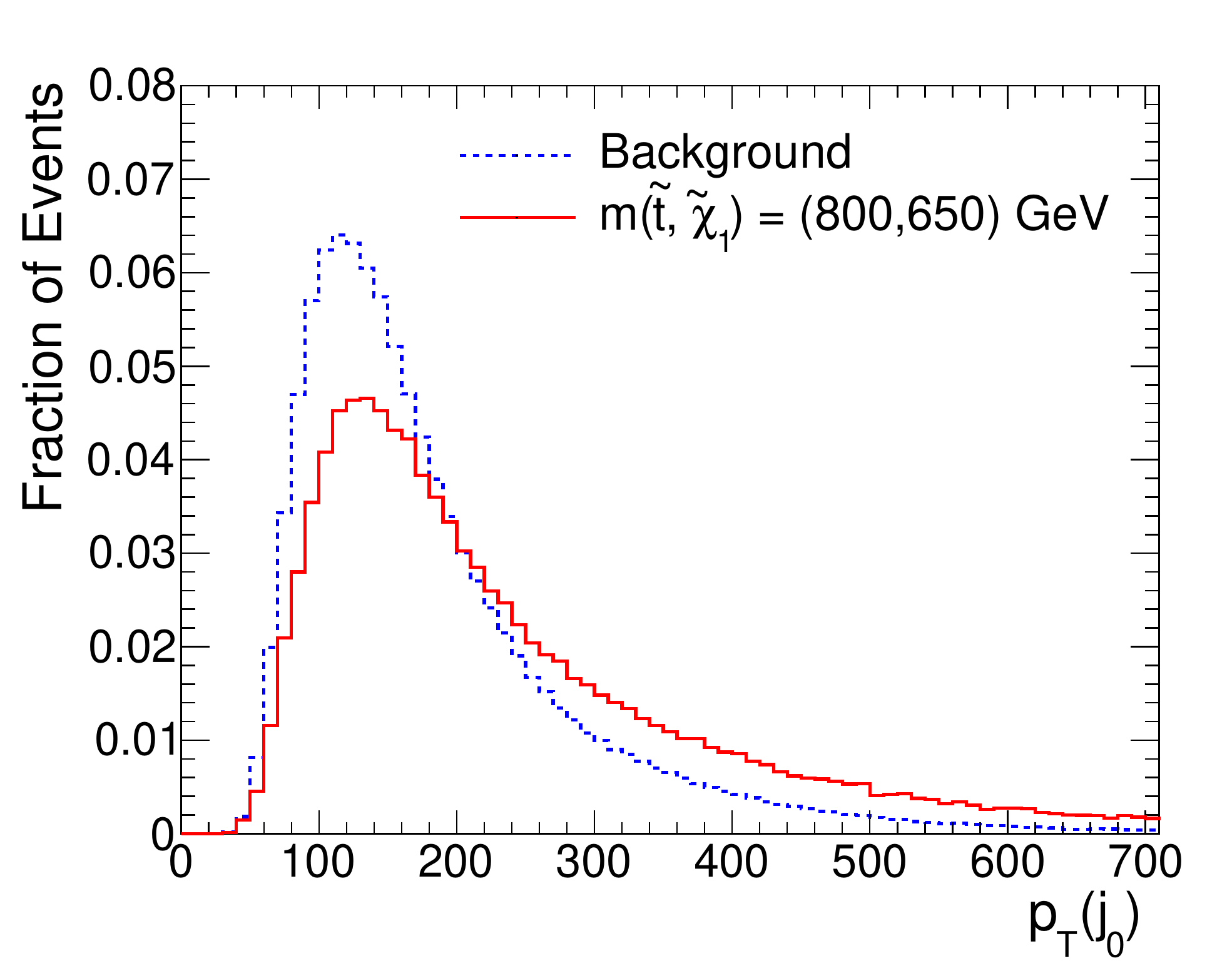}
\includegraphics[width=0.32\textwidth]{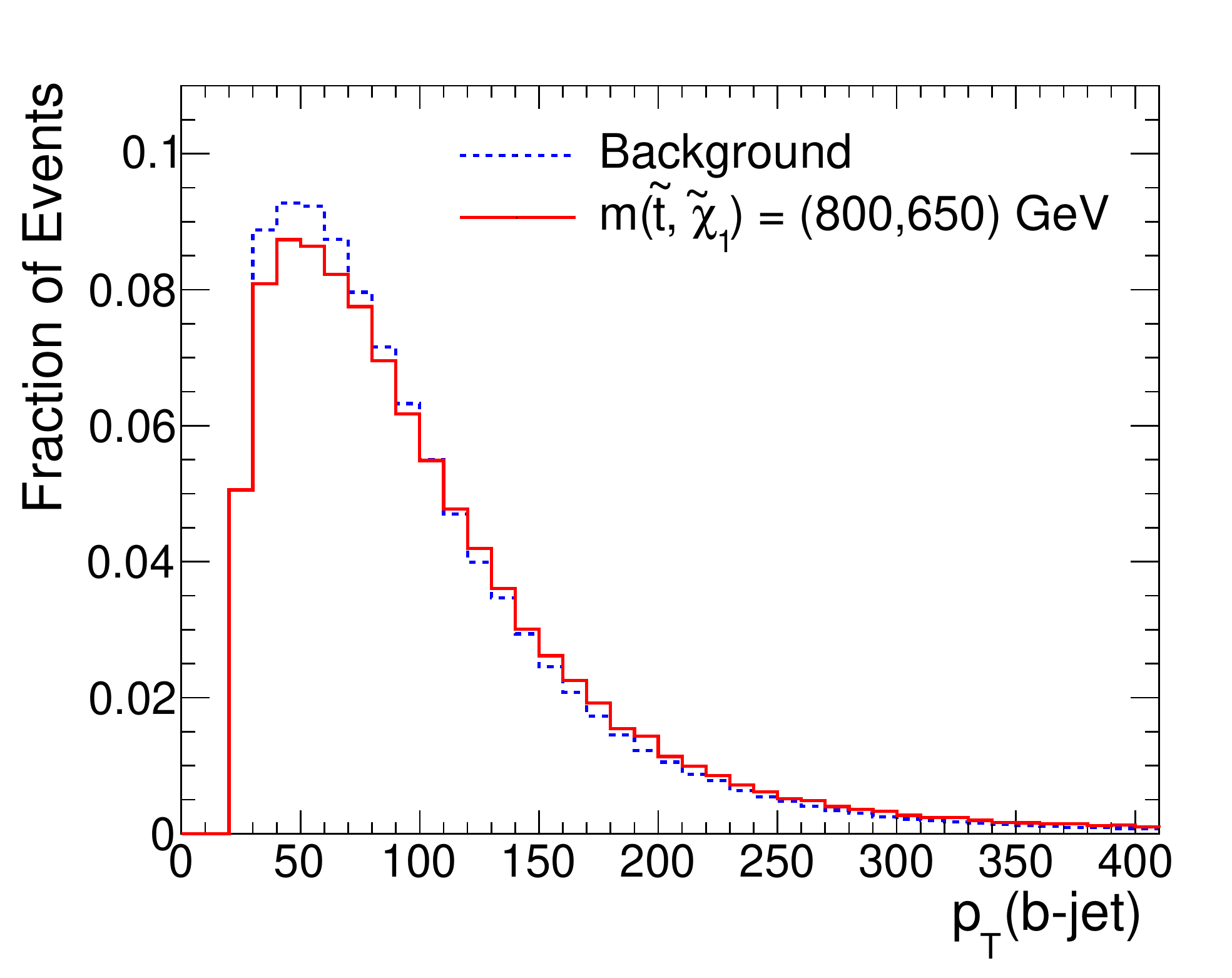}
\includegraphics[width=0.32\textwidth]{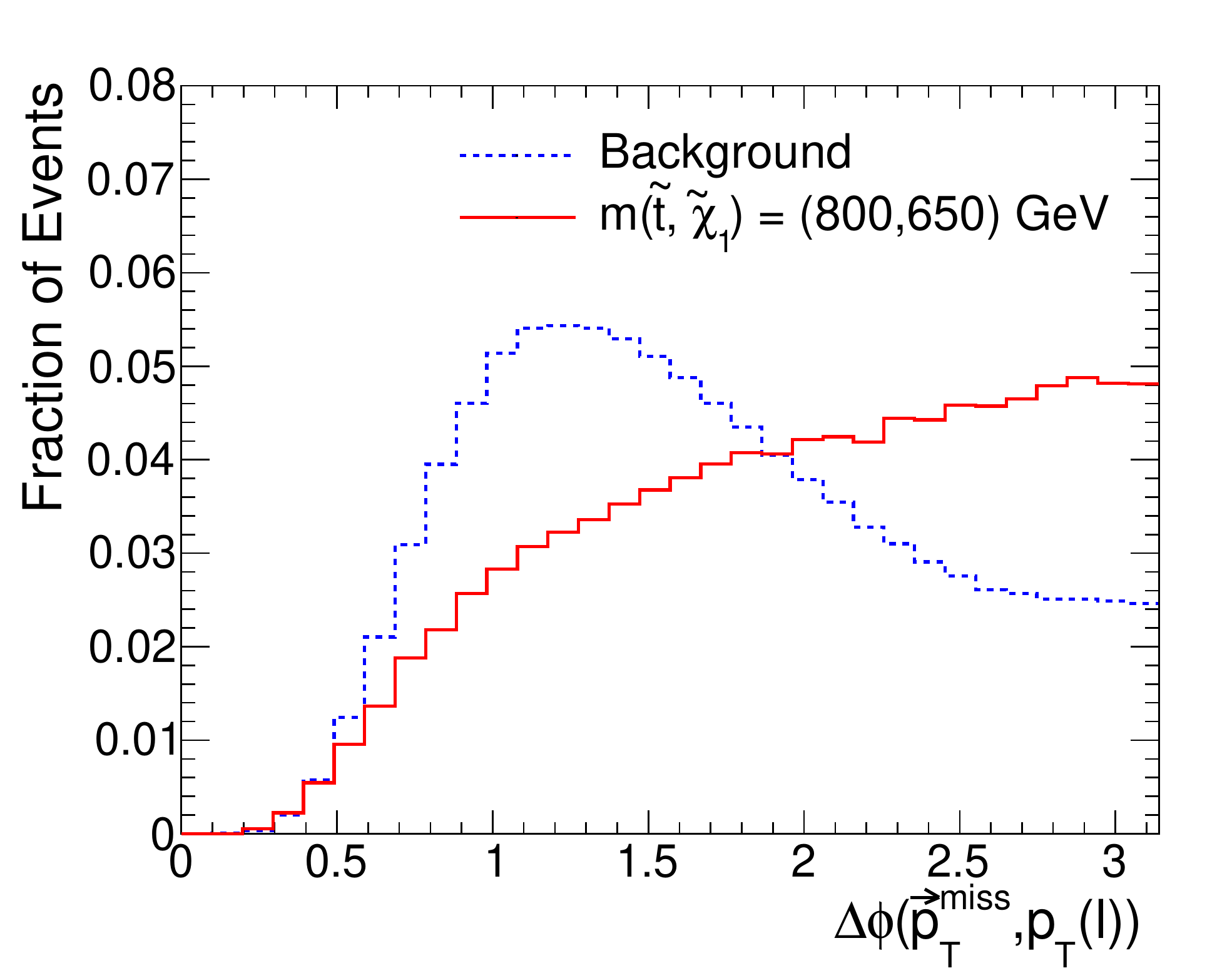}
\includegraphics[width=0.32\textwidth]{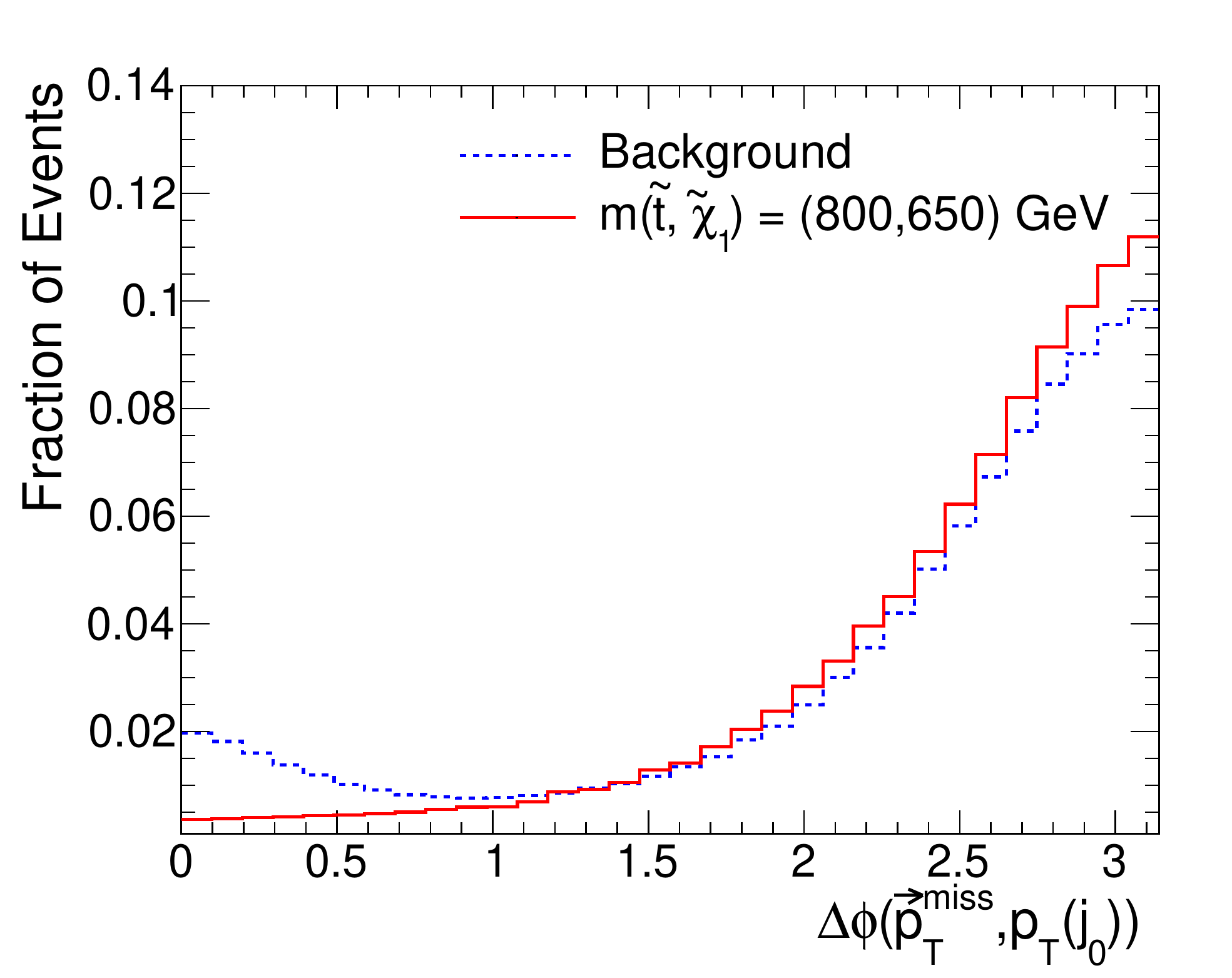}
\includegraphics[width=0.32\textwidth]{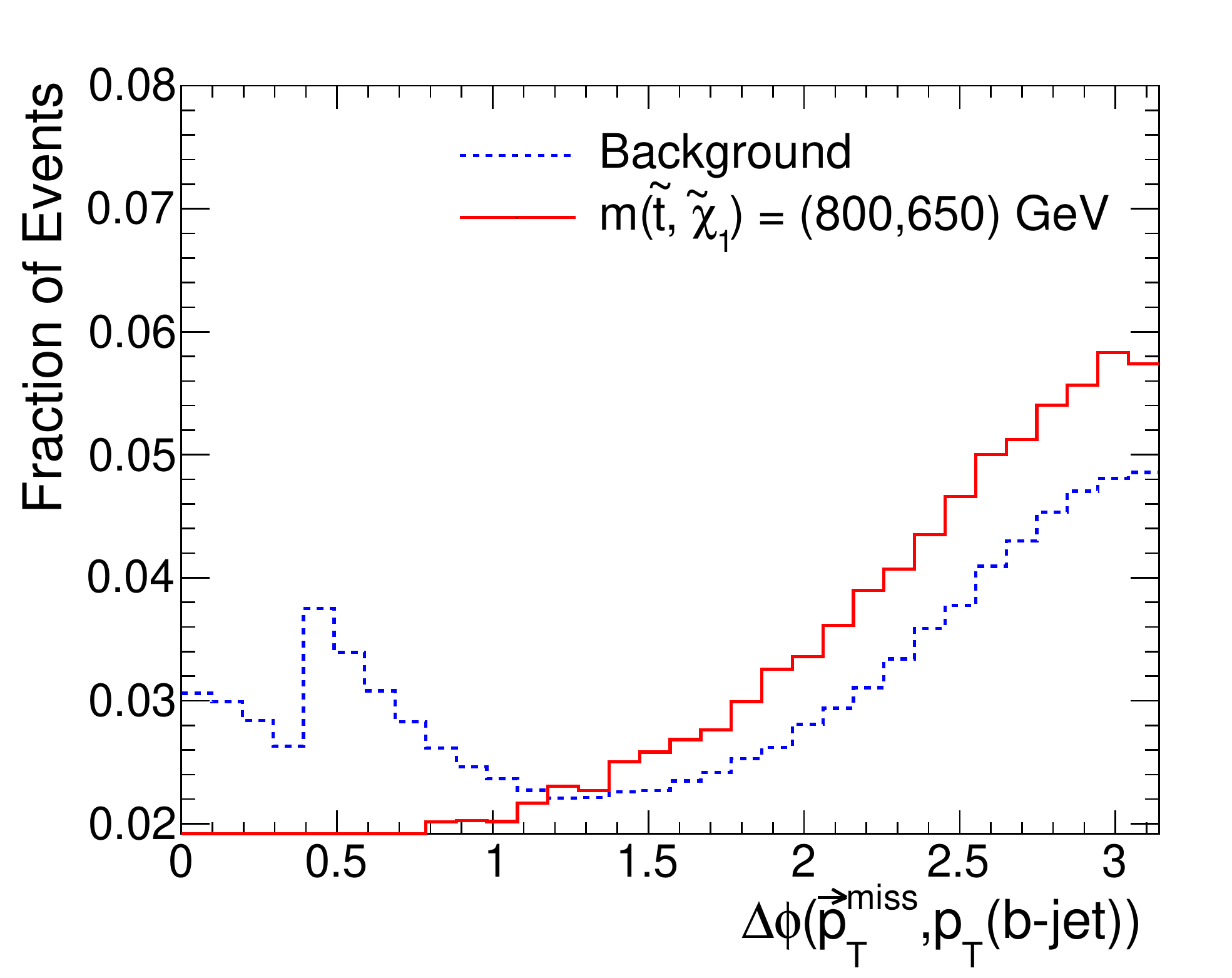}

\caption{Distributions normalized to unity for the $\textit{s800}$ signal  and backgrounds used as feature inputs in the ML training (similar variables are used for the training with the other three signal benchmarks). Here, all backgrounds have been stacked in one single histogram, according to their cross section contribution.}  
    \label{fig:features}
\end{figure}

In supervised schemes, a ML model learns from specific features that belong to different classes with the aim of classifying a new sample. In the case of high-energy physics events, these features come from physical variables obtained after event reconstruction by the experiment.
We have selected 12 of these variables from a wide set of low and high level features, knowing a priory that combining low and high level features maximizes the discriminatory power in the training \cite{features}. In our study, low level features are variables that are reconstructed directly from the detector such as  transverse momentum $p_T$, pseudo-rapidity $\eta$ and azimuthal angle $\phi$ of  leptons, jets and $b$-jets. High level features come from a combination of low level variables, like differences between azimuthal angles of particles with respect to the $\vec{p}_T^{\; miss}$, the ratio $R_M$ (eq. \ref{eq:rm}) and the transverse mass of the lepton and $E_T^{miss}$, $m_T^{lep}$ (eq. \ref{eq:mtl}).
Since there is no standardized methodology to select the most adequate features for training, we have chosen this set of variables taking into account the value of significance given after training and evaluation over the whole data set.
A list of the selected features is presented in Table \ref{tab:features}. 
Histograms with distributions normalized to unity of all of these variables are shown in Figure \ref{fig:features}. Here, all backgrounds have been stacked together and the signal shown corresponds to the case of \textit{s800} for reference. 
Some of these variables provide a good discrimination between signal and background. For instance, the $R_{M}$ and $m_{T}^{lep}$ variables have contributions at high values in the SUSY signals due to large $E_{T}^{miss}$ from the neutralinos in the SUSY benchmark points, compared to the backgrounds where the only source of $E_{T}^{miss}$ comes mainly from the neutrinos and some leptons not reconstructed correctly in the final states . We can also note differences in several low level variables such as the pseudorapidity of $\vec{p}_T^{\; miss}$, the azimuthal difference between $\vec{p}_T^{\; miss}$ and the lepton $p_T$ ($\Delta{\phi}(\vec{p}_T^{\; miss},p_T^\ell)$), and the azimuthal difference between $\vec{p}_T^{\; miss}$ and the $b$-jet transverse momentum ($\Delta \phi(\vec{p}_T^{\; miss},p_T(b\text{-jet}))$). For backgrounds, the pseudorapidity of the transverse momentum $\eta(\vec{p}_T^{\; miss})$ tend to be in the forward region in contrast with the SUSY signal which is located in the central region due to large contributions in $E_{T}^{miss}$. Furthermore, note that the $\Delta{\phi}(\vec{p}_T^{\; miss},p_T^\ell)$ variable implies that the lepton in the SUSY signal tends to be produced anti-parallel to the $\vec{p}_T^{\; miss}$, whereas for backgrounds, that difference is approximately  $\frac{\pi}{2}$. Finally, note that there is a discontinuity in the $\Delta \phi(\vec{p}_T^{\; miss},p_T(b\text{-jet}))$ distribution that comes from the pre-selection $\min \Delta\phi(j_{1,2},\vec{p}_T^{\; miss})>0.4$. 

\begin{table}[h!]
\centering
\caption{\label{tab:features} Physical variables used for training the ML classifiers. }
\begin{tabular}{cl}
\hline\hline
Input Variable & Description \\
\hline
 $E_{T}^{miss}$ & Missing transverse energy \\
 $\eta(\ell)$ & Pseudorapidity of the lepton \\
 $\eta(\vec{p}_T^{\; miss})$ & Pseudorapidity of the $\vec{p}_T^{\; miss}$ \\
 $H_T$  & Scalar sum of the $p_T$ of jets \\
 $p_{T}(\ell)$ & Transverse momentum of the lepton\\
 $p_{T}(j_0)$ & Transverse momentum of the leading jet\\
 $p_{T}(\text{b-jet})$ & Transverse momentum of the b-tagged jet\\
 $\Delta \phi(\vec{p}_T^{\; miss},p_T(\ell))$ & Azimuthal difference between $\vec{p}_T^{\; miss}$ and $p_T(\ell)$  \\
 $\Delta \phi(\vec{p}_T^{\; miss},p_T(j_0))$ & Azimuthal difference between $\vec{p}_T^{\; miss}$ and $p_T(j_0)$ \\
 $\Delta \phi(\vec{p}_T^{\; miss},p_T(b\text{-jet}))$ & Azimuthal difference between $\vec{p}_T^{\; miss}$ and $p_T(b\text{-jet}))$ \\
 $m_T^{lep}$ & Transverse mass of the lepton and $E_{T}^{miss}$ \\
 $R_{M}$     & Ratio between $E_{T}^{miss}$ and $p_{T}(j_{0})$ \\
\hline\hline
\end{tabular}
\end{table}

\subsubsection*{Data set and training}
\begin{figure}[h!]
    \centering
    \includegraphics[width=0.6\textwidth]{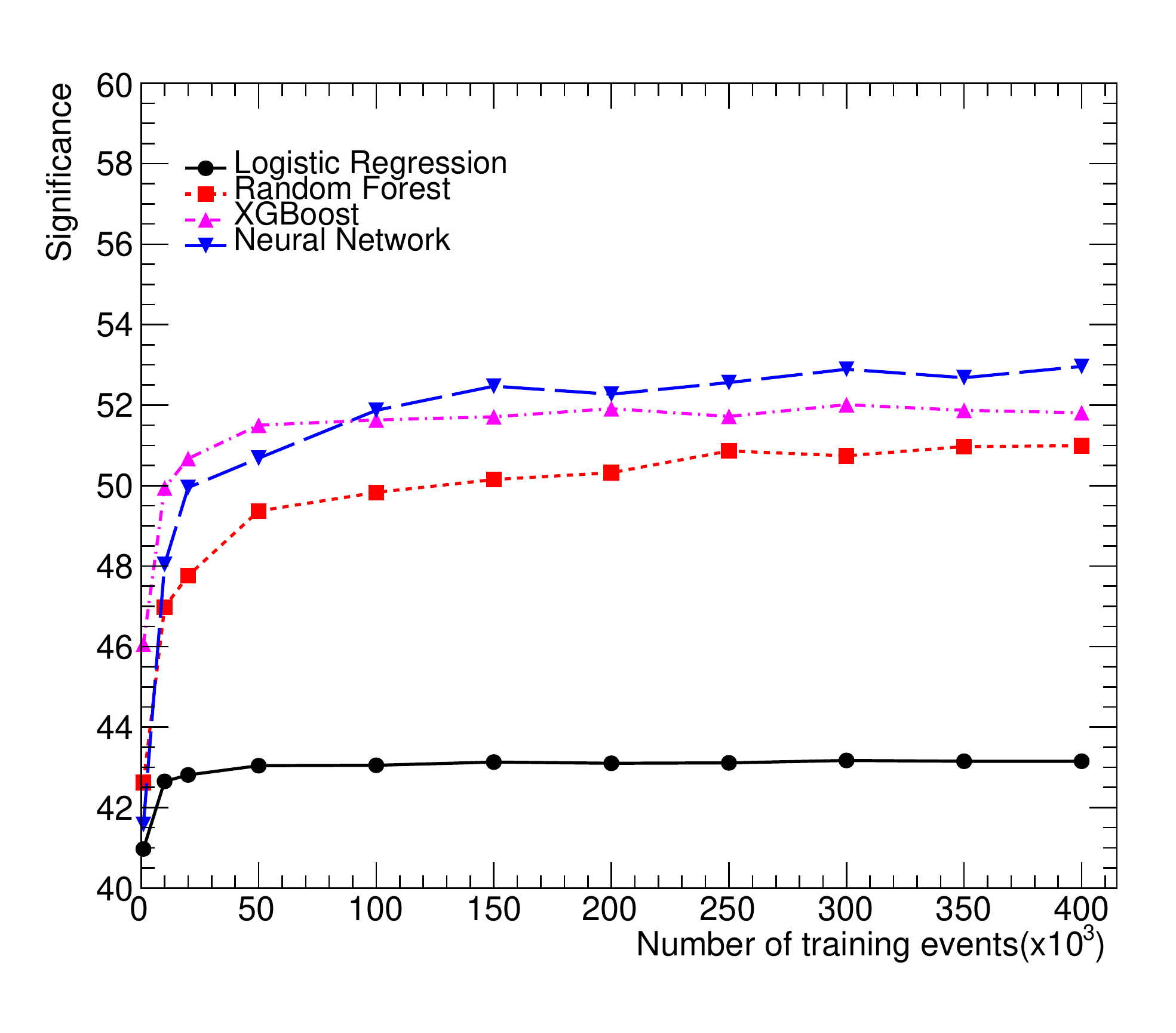}
    \caption{Significance as a function of the number of events in the training set for the \textit{s500} signal mass point and the four ML algorithms. } 
    \label{fig:sigvstraining}
\end{figure}

After a proton-proton collision, some of the final state observables  are used for  filtering  some specific event topologies by the experimental  triggers, which have  certain features that enhance some particular physics processes which are of interest for the experiment.  In our study, we are interested in events with at least one lepton (electron or muon) in the final state. Then, to target more specific events related to our final state configuration, we make the event pre-selection indicated  in  Table \ref{tab:preselections}.  Afterwards, we use the set of  variables shown in Table \ref{tab:features} for the ML training. 
In principle, it has been shown that the more data for training the better the generalization a ML algorithm can achieve \cite{deepera,scaling}. However, in this study we also consider the trade-off between the accuracy and the computational power needed to accomplish a desired performance. Consequently, due to computational constraints, we have  limited the sample size for training  to 400000  events, and use background and signal events equivalent to an integrated luminosity of 140 $fb^{-1}$, according to their cross sections, for evaluation. 
We have performed a study on how the signal significance changes when we increase the training sample, selecting as reference the \textit{s500} signal sample which has higher statistics. Results are shown in Figure \ref{fig:sigvstraining}, where we have selected the probability score that maximizes the significance for each algorithm. The significance value increases drastically  as the training  sample increases up to a value of about 50000 events for the LR, RF and XG algorithms, while the NN needs a higher training sample size (at least 100000 events) before stabilizing. The LR algorithm is the fastest to get stable values of significance with negligible gain afterwards, but with much poorer performance as compared to the other three algorithms, which reach a plateau for training samples with sizes above 250000 events. 
To ensure that we are fully sitting on the plateau of significance, we use for final training the maximal set size available of 400000 events. 
The ultimate  training, to determine maximal significance, is performed independently for each signal mass point, with 50\% of the total number of training events corresponding to the respective signal, and the    other 50\%  coming  from different sources of background combined, according to the percentage contribution (given by their cross sections), into one single background. 
From each of the total data set simulated, 90\% of the events have been used for  training and 10\% for testing.

\section{Results}
\label{sec:results}

\subsection{Results for standard cut-and-count method}

\begin{figure}[h!]
    \centering
    \includegraphics[width=0.9\textwidth]{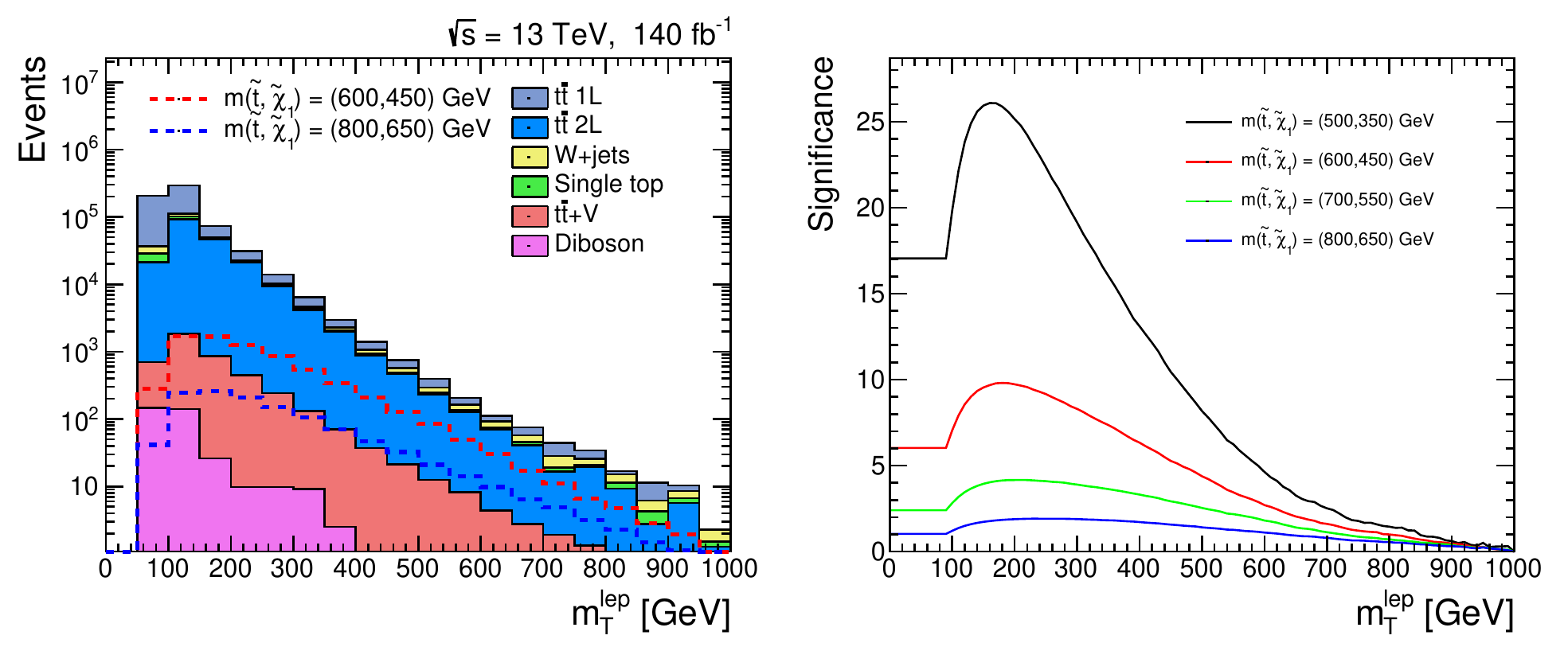}
    \includegraphics[width=0.9\textwidth]{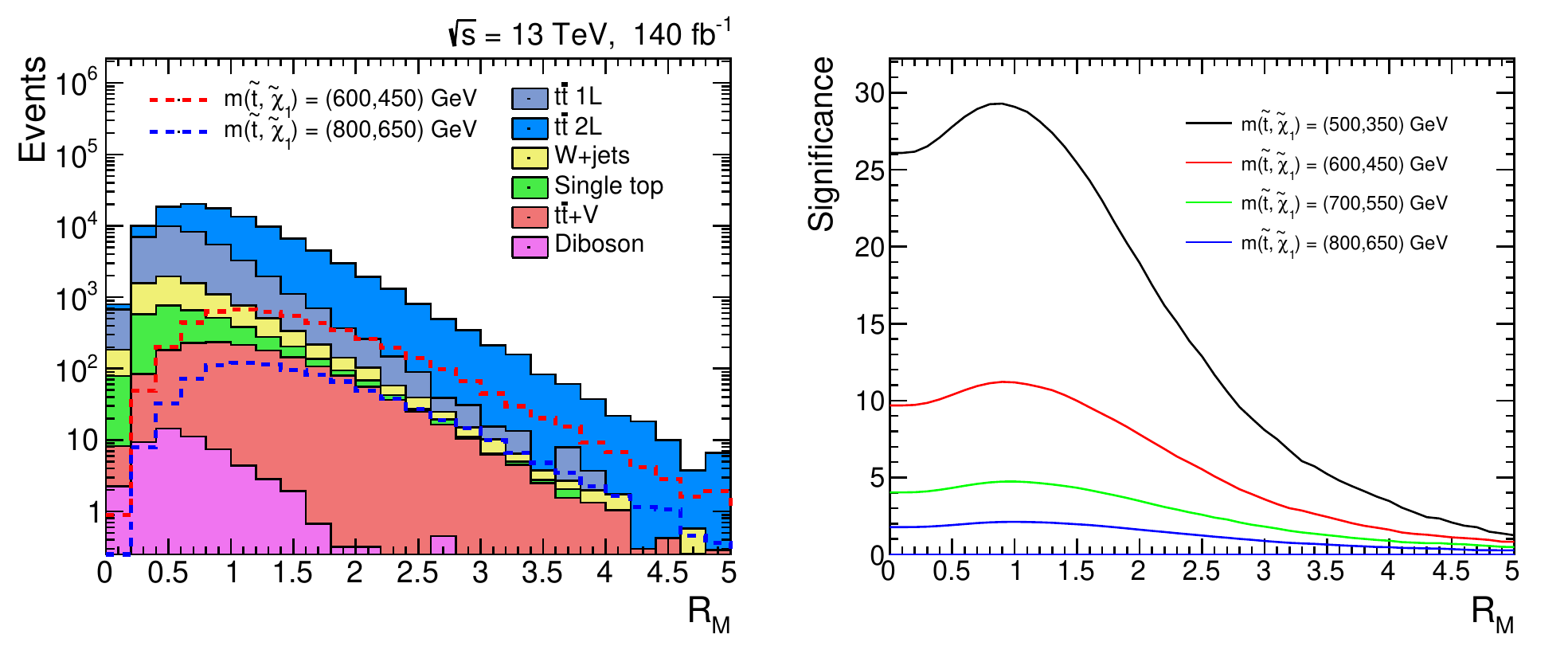}
    \includegraphics[width=0.9\textwidth]{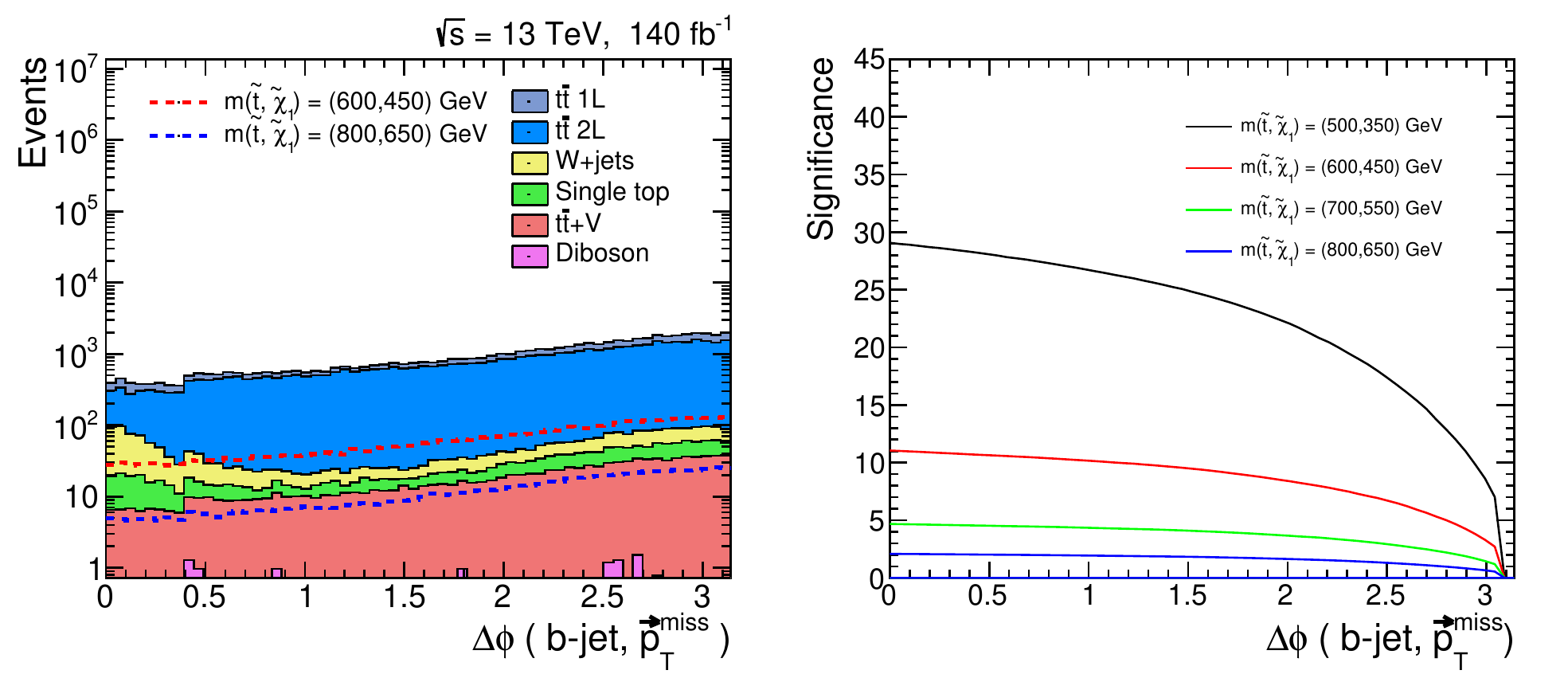}
    \caption{Optimization of significance for $m_T^{lep}$ (upper panels) after pre-selection cuts, $R_M$ (medium panels) after $m_T^{lep}>160$ GeV selection and $\Delta\phi(\text{b-jet},\vec{p}_T^{\; miss})$ (lower panels) after $m_T^{lep}>160$ GeV and $R_M>0.8$.}
    \label{fig:signrmdphi}
\end{figure}

To maximize the discrimination of stop signal from backgrounds, we have analyzed the significance variation of several physical distributions shown in Figure \ref{fig:features}. The upper panels of Figure \ref{fig:signrmdphi} indicate that the highest  significance is obtained by selecting events above  160 GeV in the $m_T^{{lep}}$ distribution. In the same way, we have found an optimal value of 0.8 for the ratio $R_M$ \cite{liantao} and this is shown in the middle panels of Figure \ref{fig:signrmdphi}.
Finally, a complementary requirement for the azimuthal angle difference between the $\vec{p}_T^{\; miss}$ and the $b$-jet $p_T$ vectors has shown to reduce significantly the $W+$jets contamination. This can be seen in the lower panels of Figure \ref{fig:signrmdphi}.
These selections are summarized in Table \ref{tab:cutsstandardanalysis}.

\begin{table}[h!]
\centering
\caption{\label{tab:cutsstandardanalysis} Event selections for signal and backgrounds within the cut-and-count methodology. }
\begin{tabular}{ll}
\hline\hline
$m_T^{{lep}}$ 	& $>160$ GeV  	\\
$E_T^{miss}/p_T^{j_0}$ 	& $>0.8$  	\\
$\Delta\phi(p_T^{\text{b-jet}},\vec{p}_T^{\; miss})$ 	& $>0.4$  	\\
\hline\hline
\end{tabular}
\end{table}

\begin{figure}[h!]
    \centering
    \includegraphics[width=0.99\textwidth]{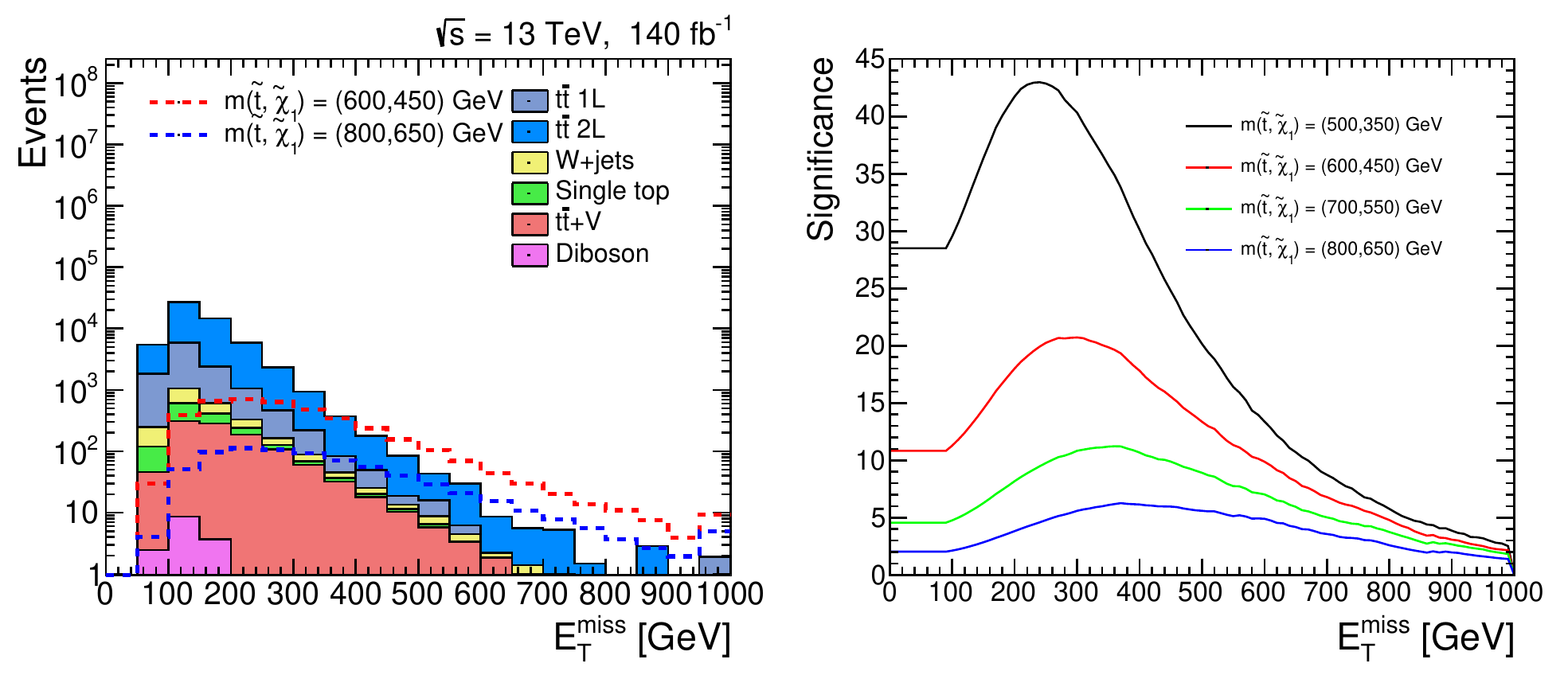}
    \caption{Event distribution for $E_T^{miss}$ (left panel) and the significance of each benchmark as a function of the $E_T^{miss}$ requirement (right panel).}
    \label{fig:metsig}
\end{figure}

After applying the previous requirements, the final $E_T^{miss}$ distribution is shown in the left panel of Figure \ref{fig:metsig} (the last bin containing events above 1000 GeV). The variation of significance with respect to the selection in $E_T^{miss}$ is shown in  Figure \ref{fig:metsig} right panel for each of the four benchmark points. A peak in this plot represents the maximum significance that can be achieved by selecting events above the respective $E_T^{miss}$ value. Table \ref{tab:sigmaxtraditional} summarizes the selections in $E_T^{miss}$ that maximize the significance for each signal benchmark. Note that these are the maximum significances  that we can  achieve by using this methodology.

\begin{table}[h!]
\centering
\caption{\label{tab:sigmaxtraditional} Maximal significance expected for each signal benchmark point after applying the indicated $E_T^{miss}$ requirement. Signal and background yields are shown for statistical reference.}

\resizebox{0.99\textwidth}{!}{
\begin{tabular}{lccrrrrrrrrc}
\hline \hline 
 & $E_T^{miss}$ & Signal       & \multicolumn{6}{c}{Backgrounds} \\ \cline{4-10} 
 & cut & Events & \multicolumn{1}{c}{$t\bar{t}$ 1L} & \multicolumn{1}{c}{$t\bar{t}$ 2L}   & \multicolumn{1}{c}{$W$+jets} & \multicolumn{1}{c}{ST} & \multicolumn{1}{c}{$t\bar{t}$+V} & \multicolumn{1}{c}{VV} & \multicolumn{1}{c}{Total} & Sig & S/B \\ \hline
\textit{s500} & 240 & 5591 & 627 & 3769 & 89 & 46 & 269 & 2 & 4802 & 43.0 & 1.2 \\
\textit{s600} & 300 & 1530 & 210  & 1270  & 40 & 19  & 133 & 1 & 1673 & 20.7 & 0.91 \\
\textit{s700} & 360 & 496  & 67  & 462  & 18  & 8  & 65  & 1 & 621  & 11.2 & 0.79 \\
\textit{s800} & 370 & 239  & 53  & 391  & 16  & 7  & 58  & 1 & 526  & 6.27 & 0.45
\\ \hline \hline 
\end{tabular} 
}
\end{table}

\clearpage
\newpage
\subsection{Results for ML algorithms}
 
In this section, we study the performance of the previously defined ML classifiers for discriminating signal from SM backgrounds and their impact on the statistical significance of a testing sample, defined in Equation \ref{eq:sig}. 

\subsubsection*{Training}

Computations for the ML training  were  performed  using  servers  with  31  Intel  Xeon  E5-2520  cores at 2.10GHz, and 32 GB of RAM memory.  Meanwhile event simulations were carried out using 24 Intel Xeon E5-2640 cores at 2.50 GHz, and 65 GB of RAM memory. All neural networks were built using the Tensorflow  machine learning software libraries \cite{tensorflow}.

\begin{figure}[h!]
\centering
    \includegraphics[width=0.49\textwidth]{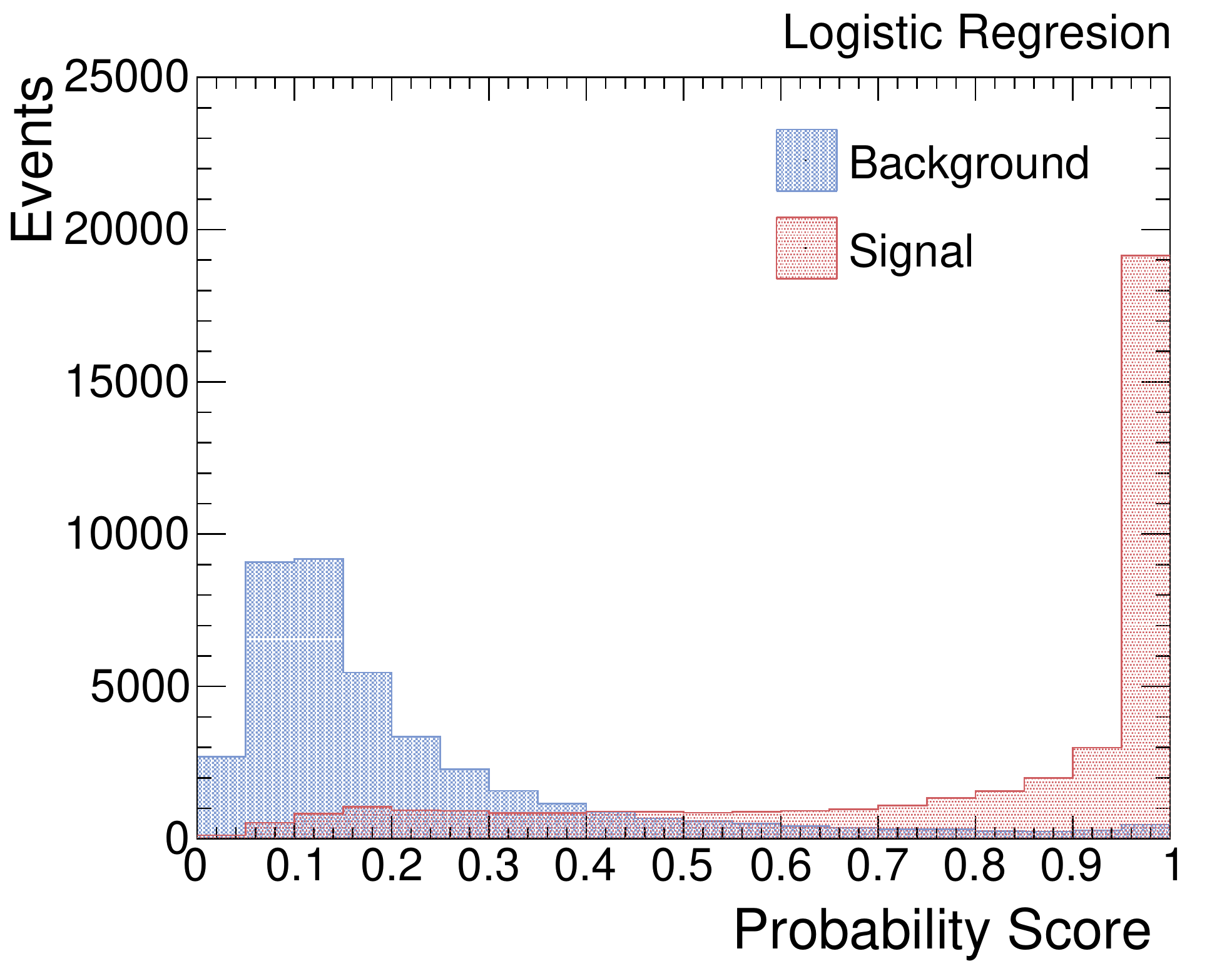}
    \includegraphics[width=0.49\textwidth]{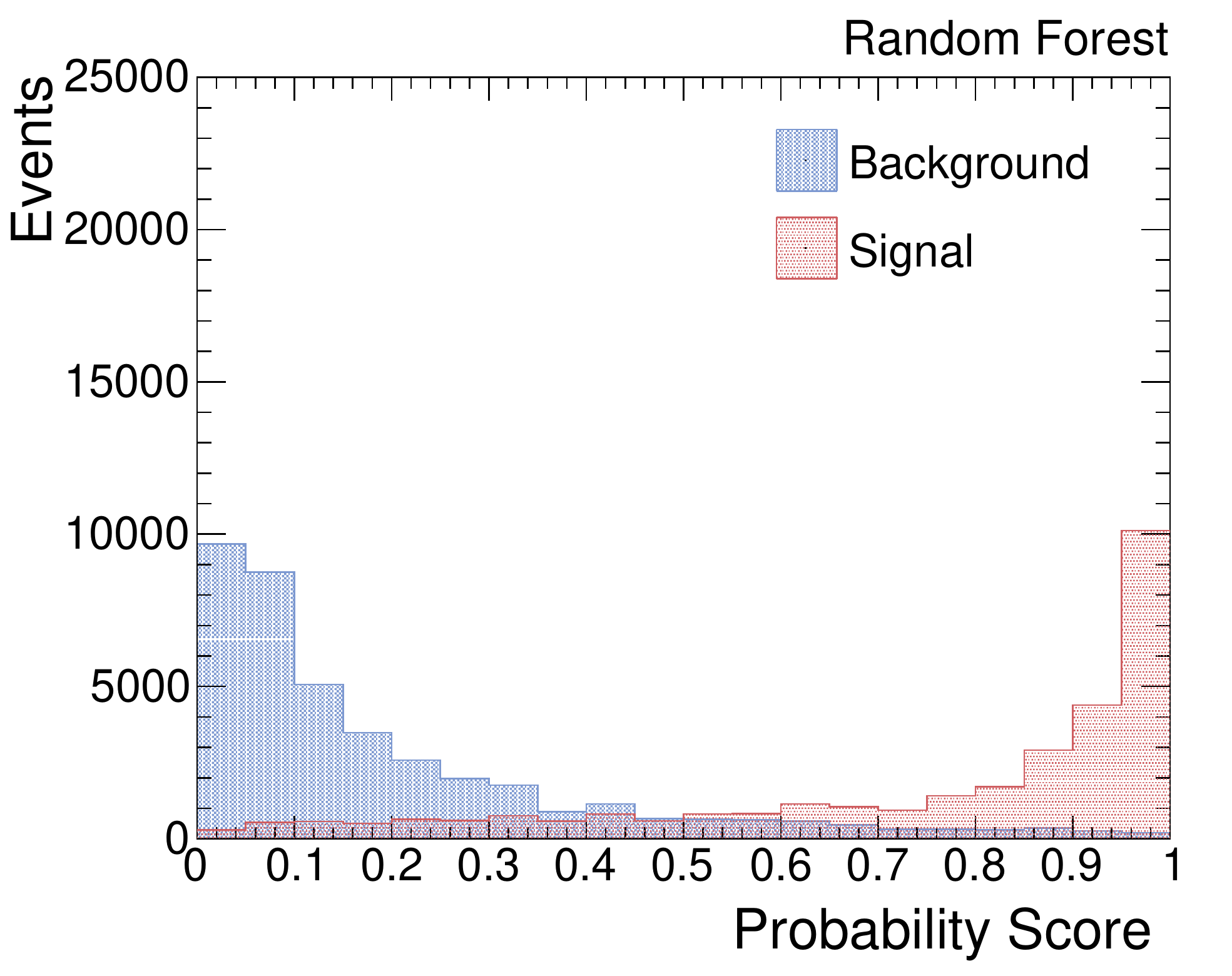}
    
    \includegraphics[width=0.49\textwidth]{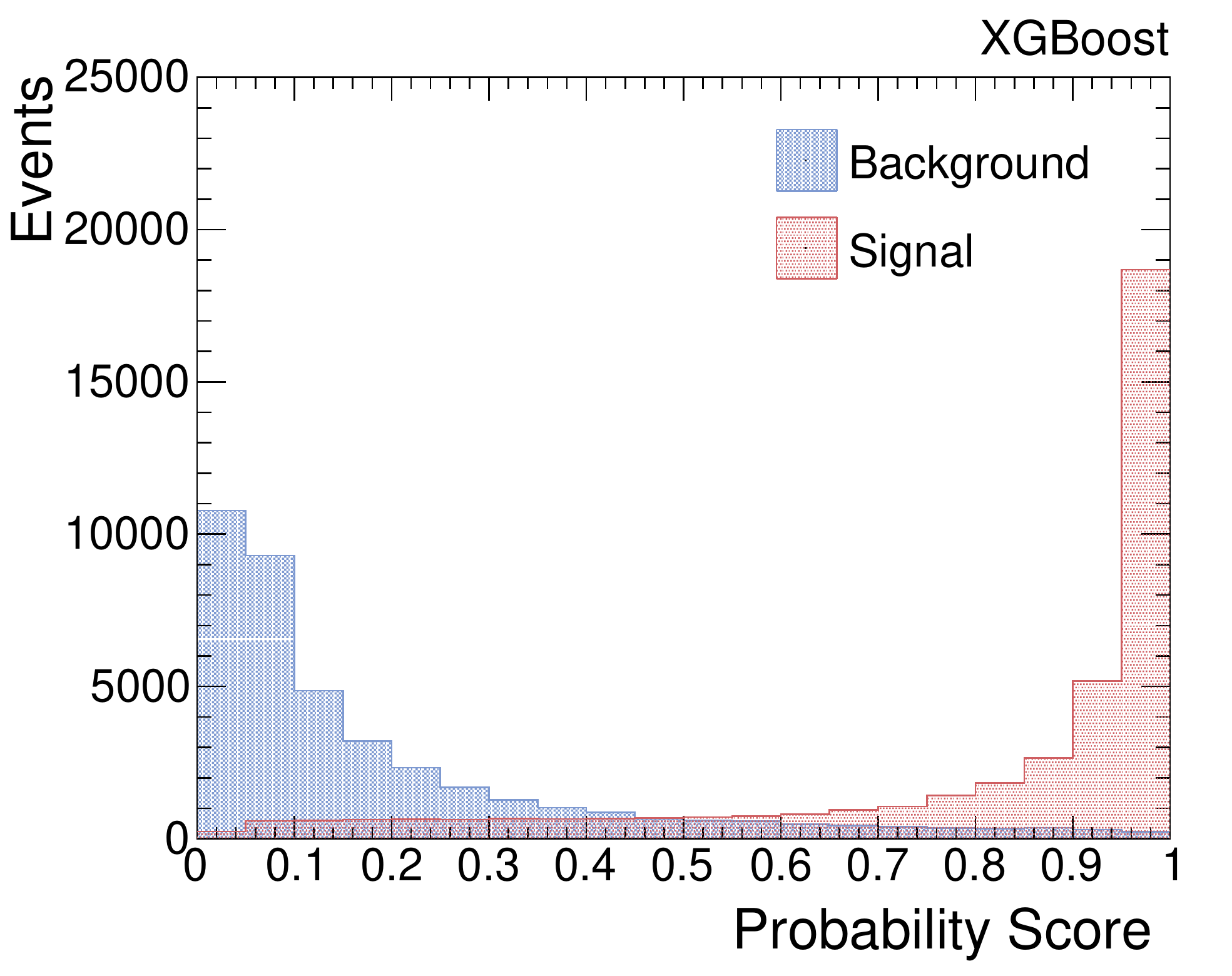}
    \includegraphics[width=0.49\textwidth]{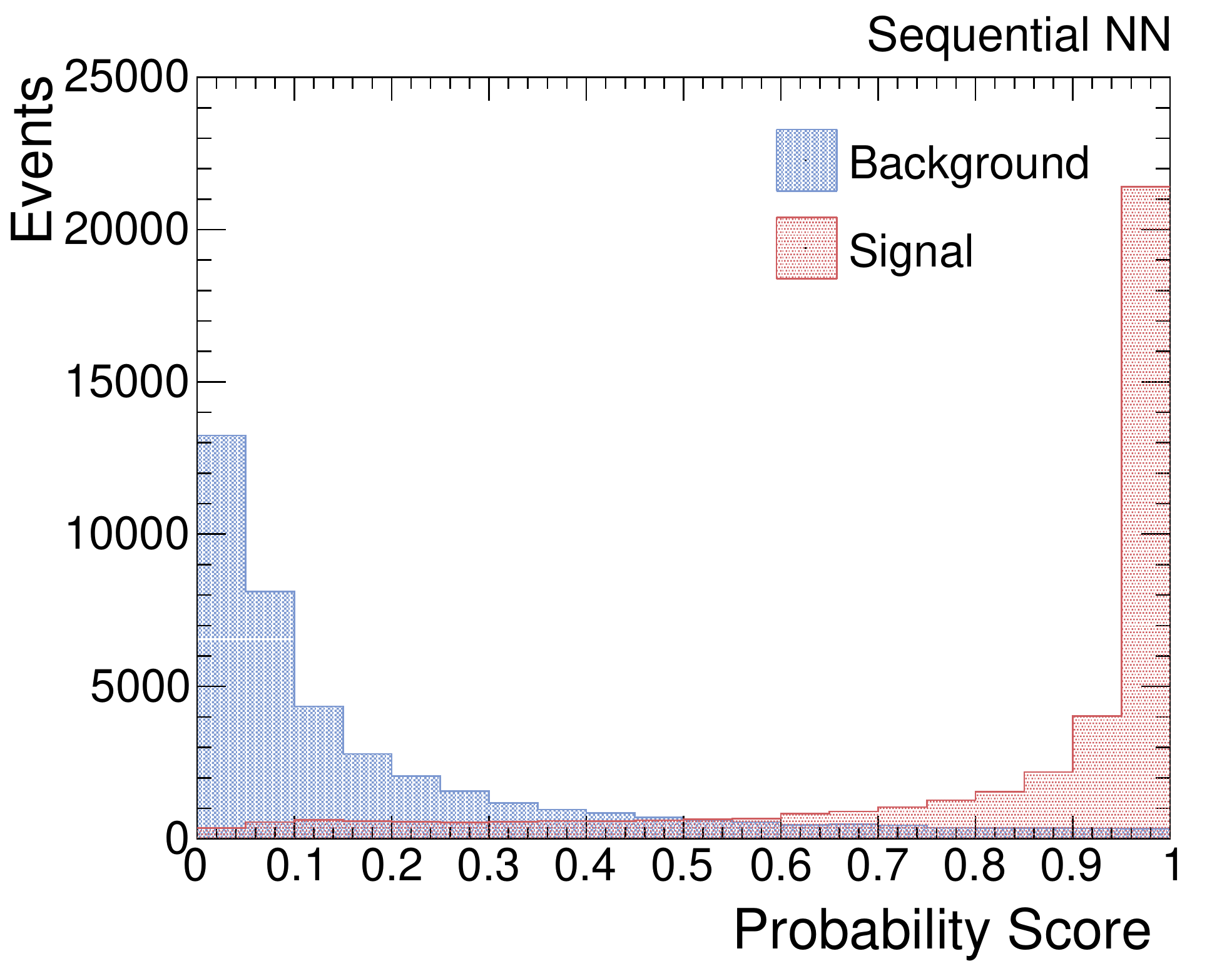}
    \caption{Probability score distributions for the testing set of events for the $m(\tilde{t}, \tilde{\chi}) = (800,650)\; \text{GeV}$, for the logistic regression method (upper left), the random forest classifier (upper right), the XGBoost classifier (lower left) and a sequential NN (lower right).}
    \label{fig:testings}
\end{figure}

We have evaluated each ML algorithm with   backgrounds and signal samples corresponding to an integrated luminosity  of 140 fb$^{-1}$. 
Distributions of the probability score for the testing set for each of these ML methods are shown in Figure \ref{fig:testings}, where the probability score  for the \textit{s800} signal mass point is shown  for reference (red shaded histogram), similar distributions are obtained for the other three signal mass points. Here, all backgrounds have been stacked and are shown in the blue shaded histogram. 
Note that the signal distribution has a peak for a probability score near unity.

\begin{figure}[h!]
    \centering
    \includegraphics[width=0.7\textwidth]{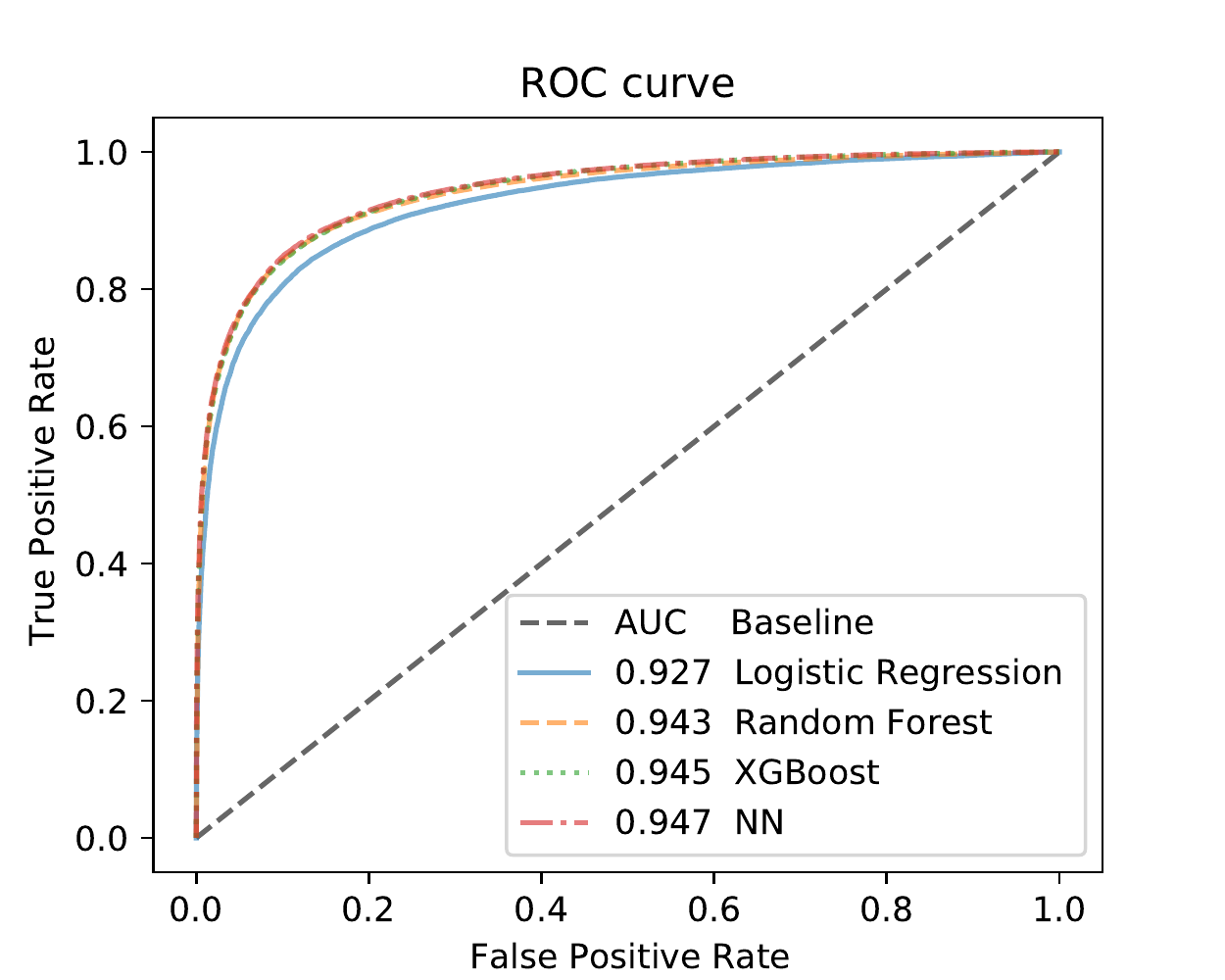}
    \caption{ROC curves and AUC values calculated for each ML method for the $m(\tilde{t}, \tilde{\chi}) = (800,650)\; \text{GeV}$.}
    \label{fig:roc}
\end{figure}

Additionally, to complement the estimation of classification performance of each ML algorithm, in Figure \ref{fig:roc} we show the receiving operating characteristic curves (ROC) \cite{roc} for the \textit{s800} signal sample. The ROC curve is intended to show the relationship between the true positive rate and false positive rate at all classification thresholds. Furthermore, by measuring the area under the ROC curve (AUC) we estimate a scale and classification-threshold invariant metric to measure the model performance. Notice that AUC is equal to 1 when all the predictions are correct (i.e. no false positives nor false negatives). In that sense, in Figure~\ref{fig:roc} we have included the AUC metric for each classifier, noting that the best classification is performed by the NN algorithm, followed closely by the XG and RF algorithms.

\subsubsection*{Evaluation}

\begin{figure}
    \centering
    \includegraphics[width=0.7\textwidth]{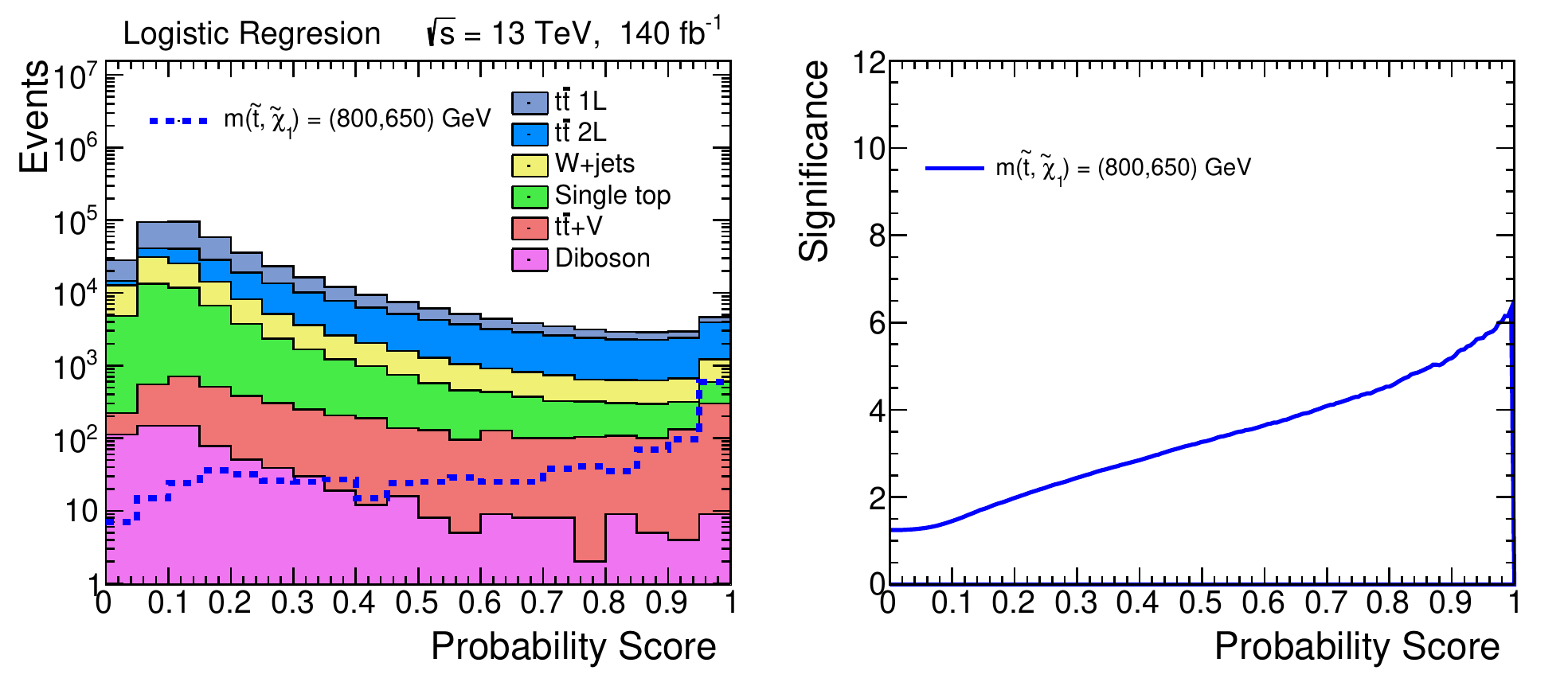}
    \includegraphics[width=0.7\textwidth]{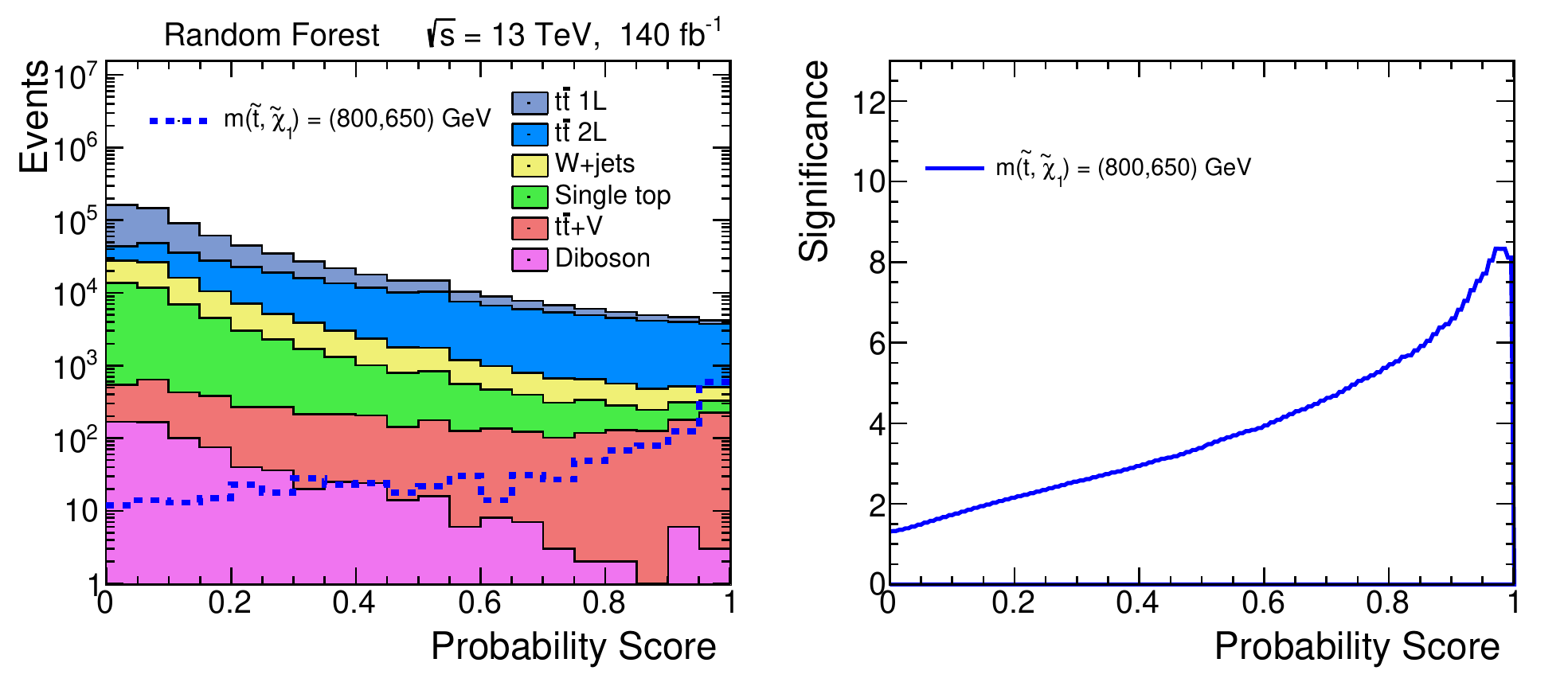}
    \includegraphics[width=0.7\textwidth]{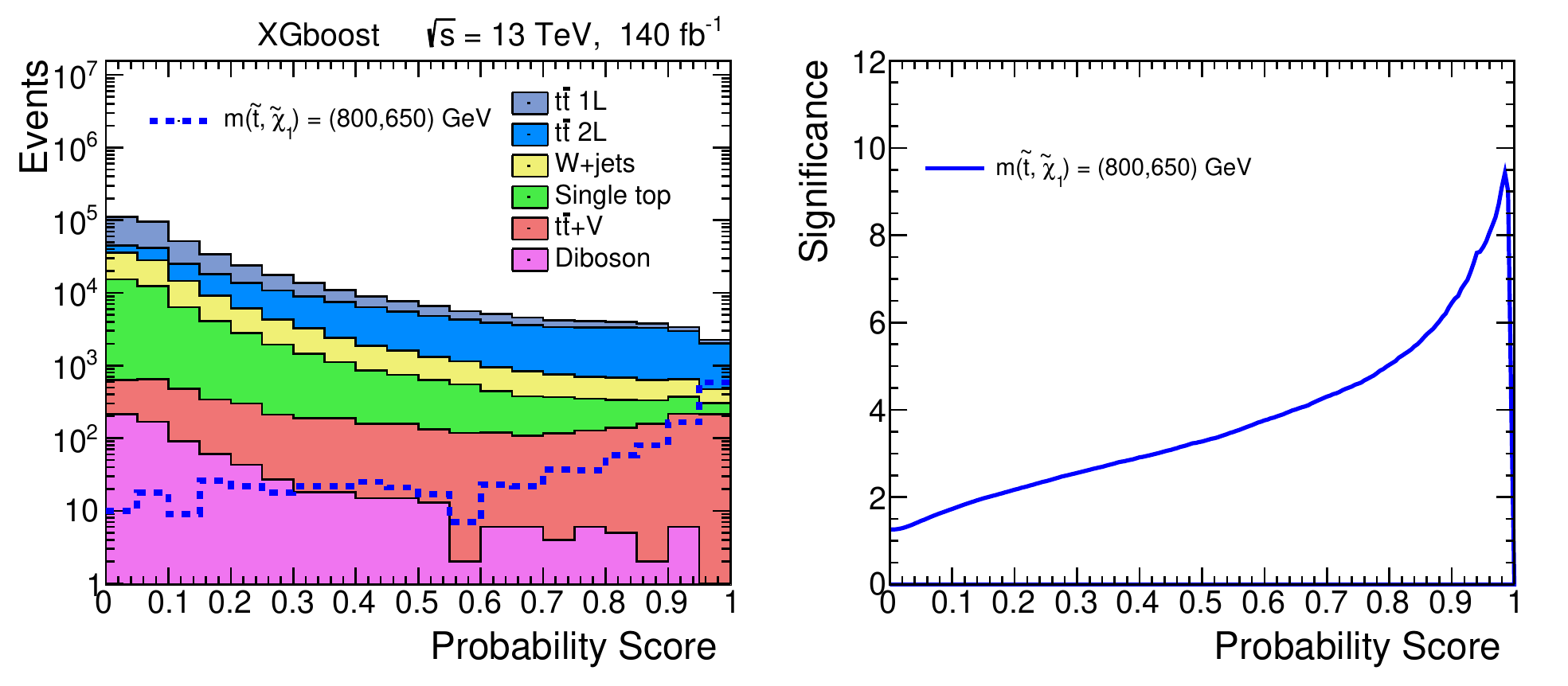}
    \includegraphics[width=0.7\textwidth]{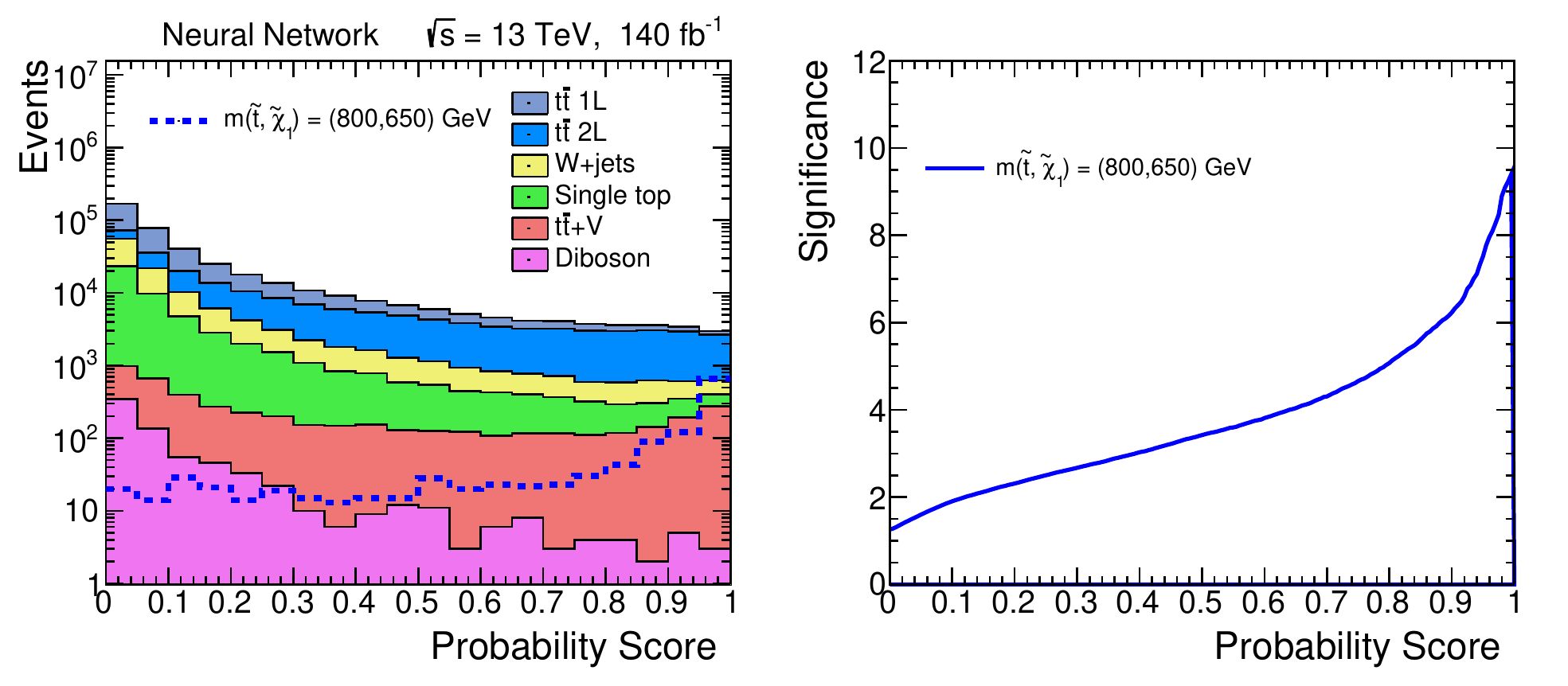}
\caption{Probability score distributions (left panels) and significance variations as a function of the probability score cut (right panels) for all ML algorithms. }
    \label{fig:sigmaxml}
\end{figure}

By using the training set, we can determine 
the threshold in the probability score of each ML algorithm that maximizes the significance of each individual  signal benchmark point. 
In the left-hand panels of Figure \ref{fig:sigmaxml} we present distributions of the probability score  for all ML algorithms for the stacked backgrounds and the \textit{s800} signal as reference. In the right-hand panels of the same Figure, the corresponding significance as a function of the  probability score is presented, where the peak of the significance is reached for a probability score close to unity. 
Since the NN and XG algorithms provide the highest significances, we list  in Tables \ref{tab:sigmaxmlnn} and \ref{tab:sigmaxmlxg}  the  thresholds used to identify the signal and background events, with the corresponding significance and the ratio of signal to background for each signal benchmark, respectively.

\begin{table}[h!]
\centering
\caption{\label{tab:sigmaxmlnn} Maximal significance expected for each signal benchmark point after applying the respective probability score threshold for the NN algorithm. Signal and detailed background yields are shown for statistical reference.}
\begin{tabular}{lccrrrrrrrcc}
\hline \hline 
 & Threshold & Signal       & \multicolumn{6}{c}{Backgrounds} \\ \cline{4-10} 
 & cut & Events & \multicolumn{1}{c}{$t\bar{t}$ 1L} & \multicolumn{1}{c}{$t\bar{t}$ 2L}   & \multicolumn{1}{c}{$W$+jets} & \multicolumn{1}{c}{ST} & \multicolumn{1}{c}{$t\bar{t}$+V} & \multicolumn{1}{c}{VV} & \multicolumn{1}{c}{Total} & Sig & $S/B$\\ \hline
\textit{s500} & 0.96 & 7745 & 740 & 4563  & 310 & 162 & 351 & 1 & 6127 & 51.9 & 1.3\\
\textit{s600} & 0.97 & 2308 & 291 & 1894  & 128 & 74  & 206 & 0 & 2593 & 25.2 & 0.89\\
\textit{s700} & 0.98 & 800  & 121 & 841  & 57  & 30  & 126 & 0 & 1175 & 13.4 & 0.68\\
\textit{s800} & 0.99 & 351  & 75  & 431   & 37  & 19  & 80  & 0 & 642  & 8.18 & 0.55\\ 
\hline \hline 
\end{tabular} 
\end{table}

\begin{table}[h!]
\centering
\caption{\label{tab:sigmaxmlxg} Maximal significance expected for each signal benchmark point after applying the respective probability score threshold for the XG algorithm. Signal and detailed background yields are shown for statistical reference.} 
\begin{tabular}{lccrrrrrrrcc}
\hline \hline 
 & Threshold & Signal       & \multicolumn{6}{c}{Backgrounds} \\ \cline{4-10} 
 & cut & Events & \multicolumn{1}{c}{$t\bar{t}$ 1L} & \multicolumn{1}{c}{$t\bar{t}$ 2L}   & \multicolumn{1}{c}{$W$+jets} & \multicolumn{1}{c}{ST} & \multicolumn{1}{c}{$t\bar{t}$+V} & \multicolumn{1}{c}{VV} & \multicolumn{1}{c}{Total} & Sig & $S/B$ \\ \hline
\textit{s500} & 0.93 & 7724 & 669 & 4988 & 262 & 147 & 372 & 1 & 6439 & 51.0 & 1.2 \\
\textit{s600} & 0.97 & 1900 & 183 & 1338 & 88  & 41  & 161 & 0 & 1811 & 24.2 & 1.0 \\
\textit{s700} & 0.98 & 687  & 91  & 621  & 39  & 24  & 87  & 0 & 862  & 13.2 & 0.80 \\
\textit{s800} & 0.98 & 380  & 109 & 708  & 51  & 27  & 103 & 0 & 998  & 7.34 & 0.38 \\ 
\hline \hline 
\end{tabular} 
\end{table}

Table \ref{tab:gain} compares the significances obtained through the standard cut-and-count methodology to the ones obtained using the ML algorithms for each signal benchmark, where the gain percentage columns correspond to the ratio between each ML algorithm significance and the ones obtained from the cut-and-count methodology.  The gain in significance is consistently better for both the NN and XG classifiers. 
While XG attains a 17-19\% gain for all benchmarks, NN provides an average 21\% for lower masses and 31\% for the \textit{s800}, being the highest gain achieved.
On the other hand, the RF algorithm provides on average a 10\% improvement, while the LR does not accomplish any enhancement over the cut and count methodology, instead reaching negative (lower than cut-and-count) performances  as low as 16\%  for the \textit{s800} benchmark.

\begin{table}[h!]
\caption{Gain in significance of the ML algorithms with respect to the traditional cut and count methodology. Accuracy in the estimate of signal significance is also provided.}
\label{tab:gain}
\centering
\resizebox{\textwidth}{!}{
\begin{tabular}{c|c|cc|cc|cc|cc}
\hline\hline
Signal  & Cut and Count   & \multicolumn{2}{|c|}{LR}     & \multicolumn{2}{|c|}{RF}     & \multicolumn{2}{|c|}{XG}  & \multicolumn{2}{|c}{NN}  \\
benchmark & Sig & Sig & Gain & Sig & Gain & Sig & Gain & Sig & Gain \\
\hline
\textit{s500} & 43.0$\pm$0.52  & 42.7$\pm$0.45  & -1\%   & 50.0$\pm$0.50 & 16\% & 51.0$\pm$0.52 & 19\% & 51.9$\pm$0.53 & 21\% \\
\textit{s600} & 20.7$\pm$0.48  & 19.3$\pm$0.39  & -7\%   & 23.1$\pm$0.42 & 12\% & 24.2$\pm$0.50 & 17\% & 25.2$\pm$0.51 & 22\% \\
\textit{s700} & 11.2$\pm$0.47  & 9.48$\pm$0.32  & -15\%  & 12.1$\pm$0.40 & 8\% & 13.2$\pm$0.47 & 18\% & 13.4$\pm$0.43 & 20\% \\
\textit{s800} & 6.27$\pm$0.39  & 5.24$\pm$0.25  & -16\%  & 6.66$\pm$0.31 & 6\% & 7.34$\pm$0.36 & 17\% & 8.18$\pm$0.45 & 31\% \\
\hline\hline
\end{tabular}
}
\end{table}

\section{Summary and conclusions}
\label{sec:conclusions}

Four ML classifiers (a Logistic Regression, a Random Forest, an XGBoost  and a Neural Network)  were  implemented  to study their potential in discriminating  hypothetical SUSY signals from their corresponding SM backgrounds. The study is based on
MC simulations of proton-proton collisions at $\sqrt{s}=13$ TeV, considering a total integrated luminosity of 140 fb$^{-1}$, and focusing on stop pair production with a
three-body decay ($\tilde{t}\rightarrow b W \tilde{\chi}^{0}_{1}$) in a compressed spectra scenario, with final states of one isolated lepton, jets, and $E_{T}^{miss}$. 
In order to target the exclusion limit of the stop-neutraliano mass phase-space  reached by ATLAS and CMS experiments with the integrated luminosity gathered in the LHC run II \cite{semileptonicatlas,semileptoniccms}, four MSSM benchmark points were considered, with stop masses between 500 GeV and 800 GeV and neutralino masses 150 GeV lower, such that the decay of the stops is produced via an off-shell top quark intermediate state.

All four ML algorithms were trained with a common set of six low-level variables (corresponding to reconstructed $p_T$ of final objects, $E_{T}^{miss}$ and angular information of lepton and $E_{T}^{miss}$) and six high-level variables (azimuthal separation of $p_{T}^{miss}$ and other final objects, transverse mass of the final lepton and $E_{T}^{miss}$, scalar addition of $p_T$ of all jets, and the ratio of $E_{T}^{miss}$ to the $p_T$ of the leading jet). For training, we have applied a selection in $E_{T}^{miss}$ greater than 90 GeV, which is much lower than ATLAS and CMS experiment requirements  of 230 GeV and 250 GeV, respectively. Therefore, we include more backgrounds, and as a consequence, our analysis is not biased towards SUSY higher mass spectra.  A study to look for the optimal size of the training set of events was executed, and performance of the ML algorithms was measured in terms of signal to background significance. Additionally, to quantify any enhancement in signal significance provided by the ML algorithms, a standard cut-and-count based analysis was implemented, following a similar approach of previous studies performed by ATLAS and CMS experiments in the channel of study.

With respect to discrimination, the conclusion is that RF, XG and NN algorithms have better performance in separating the studied SUSY compressed stop signals from backgrounds, when compared to a standard methodology based on sequential selections. While, the LR algorithm shows in average a poorer performance getting worse for higher masses. The NN and XG techniques show the highest signal significance, with XG offering an average 17\% improvement with respect to the standard cut-and-count methodology and NN showing a better enhancement, for the three lower stop masses studied and a 31\% improvement the stop mass of 800 GeV.
The close performance between NN and XG algorithms was also anticipated by the study of the ROC distributions which were very similar for both of them, as it is shown in  Figure~\ref{fig:roc}. 
It is noteworthy that the  final number of signal  events selected by both NN and XG, after applying a threshold on the probability score that maximizes significance, is over 24\% higher than the cut-and-count method, providing higher statistical power that is important when establishing any exclusion limit. 
The competing performance of the algorithm  based on Gradient boosting, as compared to NN, makes it very attractive  due to the fact that these algorithms usually require less amount of data to converge.
Furthermore, the XGBoost is much simpler to implement than the NN, as less hyper-parameters need to be optimized. For instance, in order to develop our NN we optimized the learning rate, the number of hidden layers, the number of neurons in each hidden layer and the optimizer, while for the XGBoost only the \textit{eta} (step size) and \textit{gamma}(minimum reduction to perform a further partition) were optimized.
The results obtained in the present study demonstrate the high potential and  improvement of the ML algorithms, specially NN and XG, as  efficient alternatives and consistent tools for searching for physics BSM. We have proven that they could have a substantial impact in the SUSY compressed phase space, which is experimentally more challenging in separating signal and background.

\section{Acknowledgements}

We acknowledge financial support by the ministry of science and innovation of Colombia under contract number 80740-164-2021. We also thank  the faculty of science and the department of Physics of Universidad de Los Andes, Colombia for their financial support. 


\begin{thebibliography}{99}

\bibitem{learn}
    {\it Preface to Special Issue on “Learning to Discover”,}
    \href{https://doi.org/10.1142/S0217751X20020030}{{Int. J. Mod. Phys. A, 33, (2020)}}

\bibitem{ML}
    Y. LeCun, Y. Bengio, G. Hinton,
    {\it Deep Learning,}
    \href{https://doi.org/10.1038/nature14539}{Nature, {\bf 521}, (2015) }
    
\bibitem{dnnLHC}
    D. Guest, K. Cranmer, D. Whiteson,
    {\it Deep Learning and its Application to LHC Physics,}
    \href{http://dx.doi.org/10.1146/annurev-nucl-101917-021019}{Annual Review of Nuclear and Particle Science, 68, (2018)}
    [\href{https://arxiv.org/abs/1806.11484}{arXiv:1806.11484 [hep-ex]}]
    
\bibitem{dnnLHC2}
    A. Larkoski, I. Moult, B. Nachman,
    {\it Jet Substructure at the Large Hadron Collider: A Review of Recent Advances in Theory and Machine Learning,}
    \href{http://dx.doi.org/10.1016/j.physrep.2019.11.001}{Phys. Rept., 841, (2020)}
    [\href{https://arxiv.org/abs/1709.04464}{arXiv:1709.04464 [hep-ph]}]
    
\bibitem{anomaly}
    M. Ali, N. Badrud’din, H. Abdullah, F. Kemi,
    {\it Alternate methods for anomaly detection in high-energy physics via semi-supervised learning,}
    \href{https://doi.org/10.1142/S0217751X20501316}{Int. J. Mod. Phys. A, 35, (2020)}
    
\bibitem{synchrotrons}   
     S. Appel, W. Geithner, S. Reimann, M. Sapinski, R. Singh et al. 
     {\it Application of nature-inspired optimization algorithms and machine learning for heavy-ion synchrotrons,}
     \href{https://doi.org/10.1142/S0217751X19420193}{Int. J. Mod. Phys. A, 34, (2019)}
     
\bibitem{trigger}
    T. Likhomanenko, P. Ilten, E. Khairullin, A. Rogozhnikov, A. Ustyuzhanin, M. Williams,
    {\it LHCb Topological Trigger Reoptimization,}
    \href{https://doi.org/10.1088/1742-6596/664/8/082025}{J. Phys. Conf. Ser., 664, 8, (2015)}
    [\href{https://arxiv.org/abs/1510.00572}{arXiv:1510.00572 [physics.ins-det]}]

\bibitem{calorimetry}
    D. Belayneh et al,
    {\it Calorimetry with Deep Learning: Particle Simulation and Reconstruction for Collider Physics,}
    \href{https://doi.org/10.1140/epjc/s10052-020-8251-9}{Eur. Phys. J. C, 80, 7, (2020)}
    [\href{https://arxiv.org/abs/1912.06794}{arXiv:1912.06794 [physics.ins-det]}]
         

\bibitem{dnnLHC3}
    L. de Oliveira, M. Kagan, L. Mackey, B. Nachman, A. Schwartzman,
    {\it Jet-Images -- Deep Learning Edition,}
    \href{http://dx.doi.org/10.1007/JHEP07(2016)069}{JHEP, 07, (2016)}
    [\href{https://arxiv.org/abs/1511.05190}{arXiv:1511.05190 [hep-ph]}]
    
\bibitem{dnnLHC4}
    G. Kasieczka, T. Plehn, M. Russell, T. Schell,
    {\it Deep-learning Top Taggers or The End of QCD?,}
    \href{http://dx.doi.org/10.1007/JHEP05(2017)006}{JHEP, 05, (2017)}
    [\href{https://arxiv.org/abs/1701.08784}{arXiv:1701.08784 [hep-ph]}]
    
\bibitem{mesons}
    A. M. Yasser, T. A. Nahool, M. Anwar, C. Bowerman, G. A. Yahya,
    {\it A new machine learning approach for predicting the spectra of meson bound states,}
    \href{https://doi.org/10.1142/S0218301320500925}{Int. J. Mod. Phys. E, 29, (2020)}
    
\bibitem{griddnn1} 
    M. Borisyak, F. Ratnikov, D. Derkach, A. Ustyuzhanin,
    {\it Towards automation of data quality system for CERN CMS experiment,}
    \href{http://dx.doi.org/10.1088/1742-6596/898/9/092041}{J. Phys. Conf. Ser., 9, (2017)}
    [\href{https://arxiv.org/abs/1709.08607}{arXiv:1709.08607 [physics.data-an]}]
    
\bibitem{fastdata}
    M. Grigorieva, D. Grin,
    {\it Clustering error messages produced by distributed computing infrastructure during the processing of high energy physics data,}
    \href{https://doi.org/10.1142/S0217751X21500706}{Int. J. Mod. Phys. A, 36, (2021)}
    
\bibitem{DLLHC}
    D. Guest, K. Cranmer, D. Whiteson,
    {\it Deep Learning and its Application to LHC Physics,}
    \href{http://dx.doi.org/10.1146/annurev-nucl-101917-021019}{Ann. Rev. Nucl. Part. Sci., 68, (2018)}
    [\href{https://arxiv.org/abs/1806.11484}{arXiv:1806.11484 [hep-ex]}]    
    
\bibitem{TMVA}
    A. Hoecker et al,
    {\it TMVA - Toolkit for Multivariate Data Analysis,}
    \href{https://arxiv.org/abs/physics/0703039}{arXiv:physics/0703039 [physics.data-an]}
    
\bibitem{MLjets}    
    The CMS collaboration,
    {\it Energy calibration and resolution of the CMS electromagnetic calorimeter in pp collisions at $\sqrt{s}$ = 7 TeV,}
    \href{http://dx.doi.org/10.1088/1748-0221/8/09/P09009}{JINST, 8 (2013)}
    [\href{https://arxiv.org/abs/1306.2016}{arXiv:1306.2016 [hep-ex]}]   
    
 \bibitem{griddnn2}
    V. Kuznetsov, T. Li, L. Giommi, D. Bonacorsi, T. Wildish,
    {\it Predicting dataset popularity for the CMS experiment,}
    \href{http://dx.doi.org/10.1088/1742-6596/762/1/012048}{J. Phys. Conf. Ser., 1, (2016)}
    [\href{https://arxiv.org/abs/1602.07226}{arXiv:1602.07226 [physics.data-an]}]
    
\bibitem{griddnn3}
    M. Hushchyn, P. Charpentier, A. Ustyuzhanin,
    {\it Disk storage management for LHCb based on Data Popularity estimator,}
    \href{http://dx.doi.org/10.1088/1742-6596/664/4/042026}{J. Phys. Conf. Ser., 664, 4, (2015)}
    [\href{https://arxiv.org/abs/1510.00132}{arXiv:1510.00132 [cs.DC]}]    
    
\bibitem{higgs2012atlas}
    The ATLAS collaboration,
    {\it Observation of a New Particle in the Search for the Standard Model Higgs Boson with the ATLAS Detector at the LHC,}
    \href{http://dx.doi.org/10.1016/j.physletb.2012.08.020}{Phys.  Lett. B, 716, (2012)}
    [\href{https://arxiv.org/abs/1207.7214}{arXiv:1207.7214 [hep-ex]}]

\bibitem{higgs2012cms}
    The CMS collaboration,
    {\it Observation of a new boson at a mass of 125 GeV with the CMS experiment at the LHC,}
    \href{http://dx.doi.org/10.1016/j.physletb.2012.08.021}{Phys. Lett. B, 716, (2012)}
    [\href{https://arxiv.org/abs/1207.7235}{arXiv:1207.7235 [hep-ex]}]

\bibitem{comprimidos}
    T. LeCompte, S. Martin,
    {\it Large Hadron Collider reach for supersymmetric models with compressed mass spectra,}
    \href{http://dx.doi.org/10.1103/PhysRevD.84.015004}{Phys. Rev. D, 84, (2011)}
    [\href{https://arxiv.org/abs/1105.4304}{arXiv:1105.4304 [hep-ph]}]
    
\bibitem{compressed1}  
    H. Dreiner, M. Krämer, J. Tattersall,
    {\it How low can SUSY go? Matching, monojets and compressed spectra,}
    \href{http://dx.doi.org/10.1209/0295-5075/99/61001}{EPL, 99, 6, (2012)}
    [\href{https://arxiv.org/abs/1207.1613}{arXiv:1207.1613 [hep-ph]}]
    
\bibitem{compressed2}
    J. Evans, Y. Kats, D. Shih, M. Strassler,
    {\it Toward Full LHC Coverage of Natural Supersymmetry,}
    \href{http://dx.doi.org/10.1007/JHEP07(2014)101}{JHEP, 07, (2014)}
    [\href{https://arxiv.org/abs/1310.5758}{arXiv:1310.5758 [hep-ph]}]    
    
\bibitem{neutrinos}
    J. Schechter, J.W.F. Valle,
    {\it Neutrino Masses in $SU(2)\times U(1)$ Theories,}
    \href{http://dx.doi.org/10.1103/PhysRevD.22.2227}{Phys. Rev. D {\bf 22}, 2227, (1980)}
    
\bibitem{darkmatter}
    D. Clowe, M. Bradac, A. Gonzalez, M. Markevitch, S. Randall, C. Jones, D. Zaritsky,
    {\it A direct empirical proof of the existence of dark matter,}
    \href{http://dx.doi.org/10.1086/508162}{Astrophys. J. Lett., 648, (2006)}
    [\href{https://arxiv.org/abs/astro-ph/0608407}{arXiv:astro-ph/0608407}]
    
\bibitem{darkenergy}
    E. Copeland, M. Sami, S. Tsujikawa,
    {\it Dynamics of dark energy,}
    \href{http://dx.doi.org/10.1142/S021827180600942X}{Int. J. Mod. Phys. D, 15, (2006)}
    [\href{https://arxiv.org/abs/hep-th/0603057}{arXiv:hep-th/0603057}]
    
\bibitem{cp}
    C.D. Froggatt, H. Nielsen,
    {\it Hierarchy of Quark Masses, Cabibbo Angles and CP Violation }
    \href{http://dx.doi.org/10.1016/0550-3213(79)90316-X}{Nucl. Phys. B, 147, (1979)}
    
\bibitem{hierarchy}
    A.R. Vieira, B. Hiller, M.C. Nemes, M Sampaio,
    {\it Naturalness and theoretical constraints on the Higgs boson mass,}
    \href{http://dx.doi.org/10.1007/s10773-013-1652-x}{Int. J. Theor. Phys., 52, (2013)}
    [\href{https://arxiv.org/abs/1207.4088}{arXiv:1207.4088 [hep-ph]}] 
    
\bibitem{hierarchy2}
    I. Antoniadis, N. Arkani-Hamed, S. Dimopoulos, G. Dvali,
    {\it New Dimensions at a Millimeter to a Fermi and Superstrings at a TeV,}
    \href{http://dx.doi.org/10.1016/S0370-2693(98)00860-0}{Phys. Lett. B, 436, (1998)}
    [\href{https://arxiv.org/abs/hep-ph/9804398}{arXiv:hep-ph/9804398}]

\bibitem{hierarchy3}
    N. Arkani-Hamed, S. Dimopoulos, G. Dvali,
    {\it Phenomenology, Astrophysics and Cosmology of Theories with Sub-Millimeter Dimensions and TeV Scale Quantum Gravity, }
    \href{http://dx.doi.org/10.1103/PhysRevD.59.086004}{Phys. Rev. D, {\bf 59}, (1999)}
    [\href{https://arxiv.org/abs/hep-ph/9807344}{arXiv:hep-ph/9807344}]
    
\bibitem{susytheory}
    H. Nilles,
    {\it Supersymmetry, Supergravity and Particle Physics,}
    \href{http://dx.doi.org/10.1016/0370-1573(84)90008-5}{Phys. Rept., 110, (1984)}
    
\bibitem{susytheory2}    
    H. Haber, G. Kane,
    {\it The Search for Supersymmetry: Probing Physics Beyond the Standard Model,}
    \href{http://dx.doi.org/10.1016/0370-1573(85)90051-1}{Phys. Rept., 117, (1985)}
    
\bibitem{susytheory3}
    S. Martin,
    {\it A Supersymmetry Primer,}
    \href{http://dx.doi.org/10.1142/9789812839657_0001}{Advanced Series on Directions in High Energy Physics, 21, (2010)}
    [\href{https://arxiv.org/abs/hep-ph/9709356}{arXiv:hep-ph/9709356}]     
    
\bibitem{susystates}    
    G. Farrar, P.Fayet,
    {\it Phenomenology of the Production, Decay, and Detection of New Hadronic States Associated with Supersymmetry,}
    \href{http://dx.doi.org/10.1016/0370-2693(78)90858-4}{Phys. Lett. B, 76, (1976)} 
    
\bibitem{susybroken1}
    E. Witten,
    {\it Dynamical Breaking of Supersymmetry,}
    \href{http://dx.doi.org/10.1016/0550-3213(81)90006-7}{Nucl. Phys. B, 188, (1981)}

\bibitem{susybroken2}
    E. Witten,
    {\it Constraints on Supersymmetry Breaking,}
    \href{http://dx.doi.org/10.1016/0550-3213(82)90071-2}{Nucl. Phys. B, 253, (1982)}
    
\bibitem{susybroken3}    
    L. Girardello, M.T. Grisaru,
    {\it Soft Breaking of Supersymmetry,}
    \href{http://dx.doi.org/10.1016/0550-3213(82)90512-0}{Nucl. Phys. B, 194, (1982)}  
    
\bibitem{susybroken4}
    J. Polchinski, L. Susskind,
    {\it Breaking of Supersymmetry at Intermediate-Energy,}
    \href{http://dx.doi.org/10.1103/PhysRevD.26.3661}{Phys. Rev. D, {\bf 26}, (1982)}    
    
\bibitem{susybroken5}   
    M. Dine, A. Nelson,
    {\it Dynamical Supersymmetry Breaking at Low Energies,}
    \href{http://dx.doi.org/10.1103/PhysRevD.48.1277}{Phys. Rev. D, {\bf 48}, (1993)}
    [\href{https://arxiv.org/abs/hep-ph/9303230}{arXiv:hep-ph/9303230}]

\bibitem{susybroken6}
    M. Dine, A. Nelson, Y. Shirman, 
    {\it Low-energy dynamical supersymmetry breaking simplified,}
    \href{http://dx.doi.org/10.1103/PhysRevD.51.1362}{Phys. Rev. D, {\bf 51}, (1995)}
    [\href{https://arxiv.org/abs/hep-ph/9408384}{	arXiv:hep-ph/9408384}]
    
\bibitem{strategy1}
     J. Alwall, P. Schuster, N. Toro,
     {\it Simplified Models for a First Characterization of New Physics at the LHC,}
     \href{http://dx.doi.org/10.1103/PhysRevD.79.075020}{Phys. Rev. D, {\bf 79}, (2009)}
     [\href{https://arxiv.org/abs/0810.3921}{arXiv:0810.3921 [hep-ph]}]
     
\bibitem{strategy2}
    D. Alves et al,
    {\it Simplified Models for LHC New Physics Searches,}
    \href{http://dx.doi.org/10.1088/0954-3899/39/10/105005}{J. Phys. G, 39, (2012)}
    [\href{https://arxiv.org/abs/1105.2838}{arXiv:1105.2838 [hep-ph]}]    
    
\bibitem{stops}
    C. Brust, A. Katz, S. Lawrence, R. Sundrum,
    {\it SUSY, the Third Generation and the LHC,}
    \href{http://dx.doi.org/10.1007/JHEP03(2012)103}{JHEP, 03, (2012)}
    [\href{https://arxiv.org/abs/1110.6670}{arXiv:1110.6670 [hep-ph]}]
    
\bibitem{lsp}
    G. Kane, C. Kolda, L. Roszkowski, J. Wells,
    {\it Study of Constrained Minimal Supersymmetry,}
    \href{http://dx.doi.org/10.1103/PhysRevD.49.6173}{Phys. Rev. D, {\bf{49}}, (1994)}
    [\href{https://arxiv.org/abs/hep-ph/9312272}{	arXiv:hep-ph/9312272}]
    
\bibitem{cheng} 
    Cheng, HC., Gao, C., Li, L. et al. 
    \textit{Stop Search in the Compressed Region via Semileptonic Decays.} \href{http://dx.doi.org/10.1007/JHEP05(2016)036}{J. High Energ. Phys. \textbf{2016} 36 (2016).} 
    [\href{http://arxiv.org/abs/1604.00007}{arXiv:1604.00007}]
    
\bibitem{macaluso} 
    Sebastian Macaluso, Michael Park, David Shih \& Brock Tweedie, 
    \textit{Revealing compressed stops using high-momentum recoils}, \href{http://dx.doi.org/10.1007/JHEP03(2016)151}{J. High Energ. Phys. \textbf{03} 151 (2016)} [\href{http://arxiv.org/abs/1506.07885}{arXiv:1506.07885v1}]    

\bibitem{semileptonicatlas}    
    The ATLAS collaboration, 
    {\it Search for new phenomena with top quark pairs in final states with one lepton, jets, and missing transverse momentum in pp collisions at $\sqrt{s}$= 13 TeV with the ATLAS detector,}
    \href{http://dx.doi.org/10.1007/JHEP04(2021)174}{JHEP 04 (2021) 174}
    [\href{https://arxiv.org/abs/2012.03799}{arXiv:2012.03799 [hep-ex]}]

\bibitem{semileptoniccms}   
    The CMS collaboration,
    {\it Search for direct top squark pair production in events with one lepton, jets, and missing transverse momentum at 13 TeV with the CMS experiment,}
    \href{http://dx.doi.org/10.1007/JHEP05(2020)032}{JHEP, 05, (2020)}
    [\href{https://arxiv.org/abs/1912.08887}{arXiv:1912.08887 [hep-ex]}]
    
\bibitem{dnn1}
    A. Radovic, M. Williams, D. Rousseau, M. Kagan, D. Bonacorsi, A. Himmel, A. Aurisano, K. Terao, T. Wongjirad, 
    {\it Machine learning at the energy and intensity frontiers of particle physics,}
    \href{https://doi.org/10.1038/s41586-018-0361-2}{Nature, {\bf 560}, (2018).}     

\bibitem{dnn2}
    D. Bourilkov,
    {\it Machine and Deep Learning Applications in Particle Physics,}
    \href{https://doi.org/10.1142/S0217751X19300199}{Int. J. Mod. Phys. A, 34, (2020)}
    [\href{https://arxiv.org/abs/1912.08245}{arXiv:1912.08245 [physics.data-an]}]
    
\bibitem{dnn3}
    P. Baldi, P. Sadowski, D. Whiteson,
    {\it Searching for Exotic Particles in High-Energy Physics with Deep Learning,}
    \href{https://doi.org/10.1038/ncomms5308}{Nature Commun., {\bf5}, (2014)}
    [\href{https://arxiv.org/abs/1402.4735}{arXiv:1402.4735 [hep-ph]}]
    
\bibitem{dnn4}
    M. Romao, N. F. Castro, R. Pedro,
    {\it Finding New Physics without learning about it: Anomaly Detection as a tool for Searches at Colliders, }
    \href{https://doi.org/10.1140/epjc/s10052-020-08807-w}{Eur. Phys. J. C 81, 27 (2021).}
    [\href{https://arxiv.org/abs/2006.05432}{arXiv:2006.05432 [hep-ph]}]
    
\bibitem{dnn5}
    T. Roxlo, M. Reece,
    {\it Opening the black box of neural nets: case studies in stop/top discrimination}
    [\href{https://arxiv.org/abs/1804.09278}{arXiv:1804.09278 [hep-ph]}]
    
\bibitem{dnn6}
    J. Guo, J. Li, T. Li, F. Xu, W. Zhang,
    {\it Deep learning for the R-parity violating supersymmetry searches at the LHC,}
    \href{https://doi.org/10.1103/PhysRevD.98.076017}{Phys. Rev. D, { \bf98}, (2018)}
    [\href{https://arxiv.org/abs/1805.10730}{arXiv:1805.10730 [hep-ph]}]
    
\bibitem{dnn7}
    A. Alves, 
    {\it Stacking machine learning classifiers to identify Higgs bosons at the LHC},
    \href{http://dx.doi.org/10.1088/1748-0221/12/05/T05005}{Journal of Instrumentation, {\bf 12} 05 (2017)} 
    [\href{https://arxiv.org/abs/1612.07725}{arXiv:1612.07725 [hep-ph]}] 
    
\bibitem{dnn8}
     M. Abdughani, J. Ren, L. Wu, J. Yang, J. Zhao,
     {\it Supervised deep learning in high energy phenomenology: a mini review,}
     \href{http://dx.doi.org/10.1088/0253-6102/71/8/955}{Commun. Theor. Phys., 71, 8, (2019)}
     [\href{https://arxiv.org/abs/1905.06047}{	arXiv:1905.06047 [hep-ph]}]
     
\bibitem{dnn9}
    J. Ren, L. Wu, J. Yang, J. Zhao,
    {\it Exploring supersymmetry with machine learning,}
    \href{https://doi.org/10.1016/j.nuclphysb.2019.114613}{Nucl. Phys. B, 943, (2019)}
    [\href{https://arxiv.org/abs/1708.06615}{	arXiv:1708.06615 [hep-ph]}]
    
\bibitem{dnn10}
    M. Abdughani, J. Ren, L. Wu, J. Yang,
    {\it Probing stop pair production at the LHC with graph neural networks,}
    \href{https://doi.org/10.1007/JHEP08(2019)055}{JHEP, 08, (2019)}
    [\href{https://arxiv.org/abs/1807.09088}{	arXiv:1807.09088 [hep-ph]}]
    
\bibitem{logistic}
    J. Carifio, J. Halverson, D. Krioukov, B. Nelson,
    {\it Machine Learning in the String Landscape,}
    \href{http://dx.doi.org/10.1007/JHEP09(2017)157}{JHEP, 09, (2017)}
    [\href{https://arxiv.org/abs/1707.00655}{arXiv:1707.00655 [hep-th]}]
    
\bibitem{forest}
    L. Breiman,
    {\it Random Forests,}
    \href{https://doi.org/10.1023/A:1010933404324}{Machine Learning, 45, 1, (2001)}
    
\bibitem{xgboost}
    T. Chen, C. Guestrin,
    {\it XGBoost: A Scalable Tree Boosting System,}
    \href{https://doi.org/10.1145/2939672.2939785}{Association for Computing Machinery, (2016)}
    [\href{https://arxiv.org/abs/1603.02754}{arXiv:1603.02754 [cs.LG]}]
    
\bibitem{DNN}
    P. Baldi, K. Bauer, C. Eng, P. Sadowski, D. Whiteson,
    {\it Jet Substructure Classification in High-Energy Physics with Deep Neural Networks,}
    \href{https://doi.org/10.1103/PhysRevD.93.094034}{Phys. Rev. D, {\bf9}, (2016)}
    [\href{https://arxiv.org/abs/1603.09349}{arXiv:1603.09349 [hep-ex]}]
    
\bibitem{madgraph} 
    Alwall, J., Frederix, R., Frixione, S. et al, 
    {\it The automated computation of tree-level and next-to-leading order differential cross sections, and their matching to parton shower simulations}, \href{http://dx.doi.org/10.1088/1126-6708/2002/06/029}{JHEP, 07, (2014)} 
    [\href{https://arxiv.org/abs/1405.0301}{arXiv:1405.0301}]
    
\bibitem{madgraph1}
    J. Alwall, M. Herquet, F. Maltoni, O. Mattelaer, T. Stelzer,
    {\it MadGraph 5: Going Beyond,}
    \href{http://dx.doi.org/10.1007/JHEP06(2011)128}{JHEP, 06, (2011)}
    [\href{https://arxiv.org/abs/1106.0522}{arXiv:1106.0522 [hep-ph]}]
    
\bibitem{pythia8} 
    Torbjörn Sjöstrand et al, 
    {\it An Introduction to PYTHIA 8.2,} 
    \href{http://dx.doi.org/10.1016/j.cpc.2015.01.024}{Comput. Phys. Commun., 191, (2015)}
    [\href{https://arxiv.org/abs/1410.3012}{arXiv:1410.3012[hep-ph]}]
    
\bibitem{delphes} 
    The DELPHES 3 collaboration, J. de Favereau, C. Delaere, P. Demin, A. Giammanco, V. Lemaître, A. Mertens, M. Selvaggi,
    {\it DELPHES 3: a modular framework for fast simulation of a generic collider experiment,} \href{http://dx.doi.org/10.1007/JHEP02(2014)057}{JHEP, 02, (2014)}  
    [\href{http://arxiv.org/abs/1307.6346}{arXiv:1307.6346 [hep-ex]}]
    
\bibitem{antikt} 
    Matteo Cacciari, Gavin P. Salam, Gregory Soyez, 
    \textit{The anti-$k_t$ jet clustering algorithm}   \href{http://dx.doi.org/10.1088/1126-6708/2008/04/063}{J. High Energ. Phys. 0804: 063 (2008)} [\href{https://arxiv.org/abs/0802.1189}{arXiv:0802.1189}]  
    
\bibitem{fastjet} 
    Matteo Cacciari and Gavin P. Salam, 
    \textit{Dispelling the $N63$ myth for the $k_t$ jet-finder}, \href{http://dx.doi.org/10.1016/j.physletb.2006.08.037}{Phys. Lett. \textbf{B 641} 57 (2006) } [\href{https://arxiv.org/abs/hep-ph/0512210}{arXiv:hep-ph/0512210}]
    
\bibitem{cmsdetector}
     CMS Collaboration, S. Chatrchyan,
     {\it The CMS Experiment at the CERN LHC,}
     \href{http://dx.doi.org/10.1088/1748-0221/3/08/S08004}{JINST, 3, (2008)}
    
\bibitem{softsusy}
    B. C. Allanach,
    {\it SOFTSUSY: a program for calculating supersymmetric spectra,}
    \href{http://dx.doi.org/10.1016/S0010-4655(01)00460-X}{Comput. Phys. Commun., 143, (2002)}
    [\href{https://arxiv.org/abs/hep-ph/0104145}{arXiv:hep-ph/0104145}]
    
\bibitem{crosssections}
    C. Borschensky et al,
    {\it Squark and gluino production cross sections in pp collisions at $\sqrt{s}$= 13, 14, 33 and 100 TeV,}
    \href{http://dx.doi.org/10.1140/epjc/s10052-014-3174-y}{Eur. Phys. J. C, 74, 12, (2014)}
    [\href{https://arxiv.org/abs/1407.5066}{arXiv:1407.5066 [hep-ph]}]
    
\bibitem{factors}
    L. A. Harland-Lang, A. D. Martin, P. Motylinski, R.S. Thorne,
    {\it Parton distributions in the LHC era: MMHT 2014 PDFs,}
    \href{http://dx.doi.org/10.1140/epjc/s10052-015-3397-6}{Eur. Phys. J. C, 75, 5, (2015)}
    [\href{https://arxiv.org/abs/1412.3989}{arXiv:1412.3989 [hep-ph]}]
    
\bibitem{significance1}
    R. Cousins, J. Linnemann, J. Tucker,
    {\it Evaluation of three methods for calculating statistical significance when incorporating a systematic uncertainty into a test of the background-only hypothesis for a Poisson process,}
    \href{https://doi.org/10.1016/j.nima.2008.07.086}{Nucl. Instrum. Meth. A, 595, (2008)}
    [\href{https://arxiv.org/pdf/physics/0702156.pdf}{	arXiv:physics/0702156 [physics.data-an]}]
    
\bibitem{significance2}  
    Li, T. P. , Ma, Y. Q.,
    {\it Analysis methods for results in gamma-ray astronomy,}
    \href{https://doi.org/10.1086/161295}{Astrophys. J., 272, (1983)}
 
\bibitem{significance3}  
 Cowan, G., Cranmer, K., Gross, E. et al. {\it Asymptotic formulae for likelihood-based tests of new physics}.  \href{https://doi.org/10.1140/epjc/s10052-011-1554-0}{Eur. Phys. J. C 71, 1554 (2011).}

\bibitem{uncertainties}
    ATLAS Collaboration,
    {\it Studies on top-quark Monte Carlo modelling with Sherpa and MG5aMC@NLO,}
    \href{http://cds.cern.ch/record/2261938}{ATL-PHYS-PUB-2017-007, (2017)}
    
\bibitem{uncertainties2}
    ATLAS Collaboration,
    {\it Jet energy scale measurements and their systematic uncertainties in proton-proton collisions at $\sqrt{s}$=13 TeV with the ATLAS detector,}
    \href{http://dx.doi.org/10.1103/PhysRevD.96.072002}{Phys. Rev. D, {\textbf{96}}, (2017)}
    [\href{https://arxiv.org/abs/1703.09665}{arXiv:1703.09665 [hep-ex]}]
    
\bibitem{uncertainties3}
    ATLAS Collaboration,
    {\it Performance of missing transverse momentum reconstruction with the ATLAS detector using proton-proton collisions at $\sqrt{s}$ = 13 TeV,}
    \href{http://dx.doi.org/	10.1140/epjc/s10052-018-6288-9}{Eur. Phys. J. C, 78, 2018}
    [\href{https://arxiv.org/abs/1802.08168}{arXiv:1802.08168 [hep-ex]}]
    
\bibitem{uncertainties4}
    ATLAS Collaboration,
    {\it ATLAS b-jet identification performance and efficiency measurement with $t\bar{t}$ events in pp collisions at $\sqrt{s}$=13 TeV,}
    \href{http://dx.doi.org/10.1140/epjc/s10052-019-7450-8}{Eur. Phys. J. C, 79, (2019)}
    [\href{https://arxiv.org/abs/1907.05120}{arXiv:1907.05120 [hep-ex]}]

\bibitem{liantao} 
    Haipeng An \& Lian-Tao Wang,
    \textit{Opening Up the Compressed Region of Top Squark Searches at 13 TeV LHC}, \href{http://dx.doi.org/10.1103/PhysRevLett.115.181602}{Phys. Rev. Lett. \textbf{115}, 181602, (2015)}  
    [\href{http://arxiv.org/abs/1506.00653}{arXiv:1506.00653}]     
\bibitem{logistic2}  
    A. Alves,
    {\it Stacking machine learning classifiers to identify Higgs bosons at the LHC,}
    \href{http://dx.doi.org/10.1088/1748-0221/12/05/T05005}{JINST, 12, 05, (2017)}
    [\href{https://arxiv.org/abs/1612.07725}{arXiv:1612.07725 [hep-ph]}]
    
\bibitem{esemble}
    O. Sagi  L. Rokach,
    {\it Ensemble learning: A survey,}
    Wiley Inter-disciplinary Reviews:  Data Mining and Knowledge Discovery, 8(4), e1249.
    \href{ https://doi.org/10.1002/widm.1249}{Wiley Online Library, (2018)}
    
\bibitem{forest2}
    T. Oshiro, P. Perez, J. Baranauskas,
    {\it How Many Trees in a Random Forest?,}
    \href{https://doi.org/10.1007/978-3-642-31537-4_13}{International Workshop on Machine Learning and Data Mining in Pattern Recognition, (2012)}
    
\bibitem{xgboost2} 
     L. Mason , J. Baxter , P. Bartlett , M. Frean,
     {\it Boosting Algorithms as Gradient Descent,}
     \href{https://dl.acm.org/doi/10.5555/3009657.3009730}{International Conference on Neural Information Processing, (1999)}
     
\bibitem{scikit}    
    F. Pedregosa et al,
    {\it Scikit-learn: Machine Learning in Python,}
    [\href{https://arxiv.org/abs/1201.0490}{arXiv:1201.0490 [cs.LG]}]
    
\bibitem{svm}
    Y. Zhang,
    {\it Support Vector Machine Classification Algorithm and Its Application,}
    \href{https://doi.org/10.1007/978-3-642-34041-3_27}{International Conference on Information Computing and Applications, (2012)}
    
\bibitem{regular}
    F. Salehi, E. Abbasi, B. Hassibi,
    {\it The Impact of Regularization on High-dimensional Logistic Regression,}
    [\href{https://arxiv.org/abs/1906.03761}{arXiv:1906.03761 [stat.ML]}]
    
\bibitem{gini1}    
    L. Raileanu, K. Stoffel,
    {\it Theoretical Comparison between the Gini Index and Information Gain Criteria,}
    \href{https://doi.org/10.1023/B:AMAI.0000018580.96245.c6}{Annals of Mathematics and Artificial Intelligence, {\textbf{41}}, (2004)}
    
\bibitem{rectified}
    V.Nair, G.Hinton,
    {\it Rectified linear units improve restricted boltzmann machines,}
    \href{https://dl.acm.org/doi/10.5555/3104322.3104425}{International Conference on Machine Learning, (2010)}
    
\bibitem{features}
    J. Heaton,
    {\it An Empirical Analysis of Feature Engineering for Predictive Modeling,}
    \href{https://doi.org/10.1109/SECON.2016.7506650}{IEEE Xplore, (2016)}
    [\href{https://arxiv.org/abs/1701.07852}{arXiv:1701.07852 [cs.LG]}]
    
\bibitem{deepera}   
    C. Sun, A. Shrivastava, S, Singh, A. Gupta,
    {\it Revisiting Unreasonable Effectiveness of Data in Deep Learning Era,}
    \href{https://doi.org/10.1109/ICCV.2017.97}{2017 IEEE International Conference on Computer Vision (ICCV), (2017)}
    [\href{https://arxiv.org/abs/1707.02968}{arXiv:1707.02968 [cs.CV]}]
    
\bibitem{scaling} 
    Joel Hestness et al,
    {\it Deep Learning Scaling is Predictable, Empirically,}
    [\href{https://arxiv.org/abs/1712.00409}{	arXiv:1712.00409 [cs.LG]}]
    
    
\bibitem{tensorflow}
    M. Abadi, A. Agarwal, P. Barham, E. Brevdo, Z. Chen et al,
    {\it TensorFlow: Large-Scale Machine Learning on Heterogeneous Distributed Systems,}
    [\href{https://arxiv.org/abs/1603.04467}{arXiv:1603.04467 [cs.DC]}] 
    
\bibitem{roc}
    K. Feng, H. Hong, K. Tang, J. Wang,
    {\it Decision Making with Machine Learning and ROC Curves,}
    [\href{https://arxiv.org/abs/1905.02810}{arXiv:1905.02810 [stat.ME]}]


\end{thebibliography}
\end{document}